\documentstyle[12pt,epsfig,epsf,feynarts]{article}
\newcommand{\wt}{\widetilde}
\newcommand{\ee}{e^+e^-}

\newcommand{\non}{\nonumber}
\newcommand{\nn}{\noindent}

\newcommand{\heta}{\widetilde{\theta}_f}

\renewcommand{\thefootnote}{\fnsymbol{footnote} }

\begin{document}
\begin{flushright}
MPP-2003-123\\
LPHEA-03-09\\
hep-ph/0311149
\end{flushright}

\vspace*{0.6cm}
\begin{center}
{\large{ {\bf Radiative corrections to scalar-fermion pair
production \\[0.3cm] 
in high energy e$^+$e$^-$ collisions  }}}

\vspace*{.8cm}
{\large A. Arhrib}$^{\mbox{1,2,3}}$,
{\large W. Hollik}$^{\mbox{1}}$\footnote{E-mail: 
             arhrib@mppmu.mpg.de, hollik@mppmu.mpg.de}

\vspace*{.5cm}
1: Max-Planck-Institut f\"ur Physik (Werner-Heisenberg-Institut) \\
F\"ohringer Ring 6, D--80805 Munich, Germany\\ 

\vspace{.3cm}
2: D\'epartement de Math\'ematiques, Facult\'e des Sciences et Techniques\\
B.P 416 Tanger, Morocco.\\
and\\
LPHEA, D\'epartement de Physique, Facult\'e des Sciences-Semlalia,\\
B.P. 2390 Marrakesh, Morocco.\\

\vspace{.3cm}
3: The Abdus Salam ICTP, P.O. Box 586, 34100 Trieste, Italy.

\end{center}
\vspace*{1cm}

\begin{abstract}

\vspace*{0.5cm}

\nn We study the one-loop
radiative corrections to pair production of the
supersymmetric scalar partners 
of the standard fermions 
in $\ee$ annihilation. Both electroweak and 
SUSY-QCD corrections are considered.
Applications are for
production of scalar fermions of the third generation,
$e^+e^-\to \wt{f}_i \wt{f}_j^*$ (i,j=1,2), 
$f=t, b,\tau$,  as well as for production of
scalar quarks of the first and second generation.
Effects on integrated cross sections are discussed
and also the one-loop induced
forward-backward asymmetries are studied. 
It is found that at low energy,
$\sqrt{s}\approx 500 \to 1000$ GeV, the corrections are 
dominated by the QCD contributions, 
At high energy, $\sqrt{s}\geq 2$ TeV, the electroweak box
diagrams give a substantial contribution and even dominate in some
regions of parameters space.
The purely loop-induced forward-backward asymmetry can reach values
of several per cent.

\end{abstract}
\newpage

\pagestyle{plain}
\renewcommand{\thefootnote}{\arabic{footnote} }
\setcounter{footnote}{0}

\section{Introduction}

Supersymmetry is one of the most promising extensions of the 
Standard Model (SM), and the  Minimal Supersymmetric Standard Model (MSSM)
is the popular scenario of realization of Supersymmetry \cite{susy}.
The MSSM predicts the  existence of scalar partners to all known quarks and
leptons. Since SUSY is broken, these particles can have masses 
larger than the masses of their standard partners; however, 
naturalness arguments suggest that the scale of SUSY breaking, 
and hence the masses of the SUSY particles should not exceed 
${\cal O}(1 {\rm TeV})$.
In Grand Unified SUSY models, the third generation of scalar fermions,
$\wt{t}, \wt{b}, \wt{\tau}$,
gets a special status;
due to the influence of Yukawa-coupling evolution,
the scalar fermions of the third generation are expected to be lighter 
than the scalar fermions of the first and second generations.

Scalar quarks can be produced copiously both at hadron
and lepton colliders. 
So far, the search for SUSY particles at
colliders has not been successful, and under some assumptions on the 
decay rates, one can only set lower limits on their masses.
CDF excludes scalar quarks of the first and second generations
with masses lower than ${\cal O}(250$ GeV) \cite{CDF},
while for scalar top and scalar bottom the bounds are lower.
LEP experiments also searched for scalar fermions,
depending on assumptions on the 
Lightest Supersymmetric Particle (LSP); the results can be 
summarized as follows \cite{LEP}:
\begin{itemize}
\item from $e^+e^- \to \wt{t}_1\wt{t}_1$, 
and assuming that the splitting $m_{\widetilde{t}_1}-m_{LSP} 
\geq 10 $ GeV :\\
i) if the stop decays as
$\wt{t}_1 \to c \wt{\chi}_1^0$, the lower limit 
on the $\wt{t}$ mass is
97.6 (95.7) GeV for stop mixing $\wt\theta_{{t}}=0$ 
($\wt\theta_{{t}}=0.98$).\\
ii) if the stop decays as
$\widetilde{t}_1 \to b l \widetilde{\nu}$, $l=e, \mu, \tau$, 
the lower limit is 
96 (92.6) GeV for stop mixing $\wt{\theta}_{{t}}=0$ 
($\wt\theta_{{t}}=0.98$).\\
iii) if the stop decays as
$\widetilde{t}_1 \to b \tau \widetilde{\nu}_\tau $ with 100\% branching ratio, 
the lower limit is 95.5 (91.5) GeV for stop mixing 
$\wt{\theta}_{t}=0$ ($\wt\theta_{{t}}=0.98$).
\item from $e^+e^- \to \wt{b}_1\wt{b}_1$ followed by
the decay $\wt{b}_1 \to b  \wt{\chi}_1^0 $
and assuming for the splitting $m_{\wt{b}_1}-m_{LSP} 
\geq 10 $ GeV, the lower limit on the $\wt{b}$ mass is 96.9 (85.1) GeV for 
sbottom mixing $\wt{\theta}_{b}=0$ ($\wt\theta_{{b}}=1.17$).
\end{itemize}
Note that a mixing scenario with $\wt{\theta}_{t}=0.98$ 
($\wt{\theta}_{b}=1.17$) corresponds to the case 
where the light scalar top decouples from the $Z$ boson, with
the $Z\wt{t}_1\wt{t}_1^*$ coupling $\approx 0$
(or the light scalar bottom decouples from $Z$ boson, with
the $Z\wt{b}_1\wt{b}_1^*$ coupling $\approx 0$, respectively).

Higher-energy hadron and $\ee$ colliders of the next generation
will be required to sweep the entire mass range, up to 
${\cal O}$ (1TeV), for the supersymmetric particles.
If SUSY particles would be detected at hadron colliders,
their properties can be studied with high accuracy at a 
high-energy linear $e^+ e^-$ collider~\cite{LC}. It has been shown that 
the expected experimental precision for slepton-mass measurements 
is of the order of 100 MeV \cite{LC}.
It is thus mandatory to incorporate effects beyond leading order
into the theoretical predictions in order to match the 
experimental accuracy.
In particular,
QCD corrections to squark-pair production have to be included. 
The next-to-leading order corrections to squark-pair 
production at proton colliders have been studied in \cite{DESY} and found
to increase the cross section. 

For $\ee$ machines,
scalar-fermion production has been addressed in 
several studies and shown to be promising for
precision analysis of sfermion
properties with mass and
mixing-angle reconstructions~\cite{LC,S.SU}. 
QCD corrections to squark-pair production 
were  shown, a decade ago, to be large \cite{DH,ACD,helmut1},
and the gluino contributions~\cite{ACD,helmut1} 
were found to be rather small.
Gluon emission from scalar quarks
has been studied in~\cite{zerwas}.
On the electroweak side,
for squark and slepton production in $e^+e^-$ annihilation
the leading and subleading electroweak 
Sudakov logarithms were investigated~\cite{claudio1, claudio2}
and found to be large at high energy. 
Recently, the full one-loop radiative corrections to 
the production of scalar muons and
scalar electrons have been presented in~\cite{freitas1}, and
in refs~\cite{freitas1,freitas2}, it was shown that 
near threshold important corrections to the cross section arise from 
Coulomb corrections and finite slepton widths.
A few years ago, a part of the Yukawa corrections
(the one associated with vertex corrections) to the third generation
of scalar fermions was considered in~\cite{helmut2}.

In this paper we provide the complete result 
for the electroweak corrections to scalar-fermion
pair production in $\ee$ annihilation, including 
self energies, vertex corrections,
box diagrams,
and real photon emission, and discuss their effects  
in combination with the QCD corrections.
As applications, we
focus on two types of observables: total cross sections and
forward--backward asymmetries, the latter one being purely loop-induced.

The paper is organized as follows. In the next section, we will first set the
notations and give the 
tree-level results.
Section 3 outlines
the calculation and  
the renormalization scheme.
In section 4 we will discuss the effects of radiative corrections
for various types of sfermions, 
with a short conclusion in section 5.

\renewcommand{\theequation}{2.\arabic{equation}}
\setcounter{equation}{0}
\section{Notation and tree-level results}

First we summarize the MSSM parameters entering our analysis,
with particular attention to the sfermion sector. 
In the MSSM, the sfermion sector is specified by the mass matrix in 
the basis $(\tilde{f}_L^{},\tilde{f}_R^{})$. In terms of 
the scalar masses $\widetilde{M}_L$, $\widetilde{M}_R$, 
the Higgs-Higgsino mass parameter $\mu$, and
the soft SUSY-breaking trilinear couplings $A_f$, the sfermion 
squared-mass matrices are given by~\cite{gun} 
\begin{equation}
{\cal M}^2_{\tilde{f}}= 
     \left( \begin{array}{cc} 
                m_f^2 + m_{LL}^2 & m_{LR} m_f \\
                m_{LR} m_f     & m_f^2 + m_{RR}^2
            \end{array} \right) \label{eq1}
\end{equation}
with
\begin{eqnarray}
  m_{LL}^2 &=& \widetilde{M}_{L}^2 
    + m_Z^2\cos 2\beta\,( I_3^f - Q_f s_W^2 ) ,\\
  m_{RR}^2 & = & \widetilde{M}_{R}^2  
                  + m_Z^2 \cos 2\beta\, Q_f s_W^2 , \label{eq:c} \\[2mm]
  m_{LR}    &=& 
  A_f - \mu\ (\tan\beta)^{-2I_3^f} \,\, . \label{eq:d}
\end{eqnarray}
$I_3^f=\pm 1/2$ and $Q_f$ are the weak isospin and the electric charge of 
the sfermion $\tilde{f}$, and $\tan\beta = v_2/v_1$ with 
the vacuum expectation values of the Higgs fields,
$v_1$ and $v_2$.

The hermitian matrix (\ref{eq1}) is diagonalized by 
a unitarity matrix 
$R^{{\tilde{f}}}$, which rotates the current eigenstates,
${\tilde{f}}_L$ and ${\tilde{f}}_R$, into the mass
eigenstates $\tilde{f}_1$ and $\tilde{f}_2$ as follows,
\begin{equation}
\left( \begin{array}{c} 
                {\tilde{f}}_1 \\
                {\tilde{f}}_2
\end{array} \right) = R^{{\tilde{f}}} 
\left( \begin{array}{c} 
{\tilde{f}}_L \\
{\tilde{f}}_R
\end{array} \right) = 
\left( \begin{array}{cc} 
 \cos{\wt{\theta}_f}  & 
 \sin{\wt{\theta}_f}   \\
- \sin\wt{\theta}_f &
 \cos\wt{\theta}_f
            \end{array} \right) 
\left( \begin{array}{c} 
                {\tilde{f}}_L \\
                {\tilde{f}}_R 
\end{array} \right) \, ,
\label{eqe}
\end{equation}
yielding the physical mass eigenvalues, with the convention
$m_{{\tilde{f}}_{1}}< m_{{\tilde{f}}_{2}}$,  
\begin{eqnarray}
m_{{\tilde{f}}_{1,2}}^2 &=& \frac{1}{2}(2 m_f^2 + m_{LL}^2 + m_{RR}^2 
\mp \sqrt{ (m_{LL}^2 - m_{RR}^2)^2 + 4 m_{LR}^2 m_f^2 })\ . 
\label{mass} 
\end{eqnarray}
The mixing angle $\wt{\theta}_f$ obeys the relation  
\begin{eqnarray}
\tan 2\wt{\theta}_f =\frac{2 m_{LR} m_f}{m_{LL}^2 -m_{RR}^2 } \ \ . \label{mixing}
\end{eqnarray} 
Hence, 
for the  case of the supersymmetric partners of the light fermions, 
$L$--$R$ mixing can be neglected.
However, mixing between top squarks can be
sizable and allows one of the two mass eigenstates
to be lighter than the top quark. Bottom-squark and $\tau$-slepton
mixing can also be significant if $\tan\beta$ is large.

The interaction of the neutral gauge bosons $\gamma$ and $Z$ 
with the sfermion-mass eigenstates is described  by the Lagrangian
\begin{eqnarray}
 {\cal L} = & & -i e A^\mu \sum_{i=1,2} Q_f \tilde{f}_i^* 
\stackrel{\leftrightarrow}{\partial}_\mu
 \tilde{f}_i + i Z^\mu \sum_{i,j=1,2} g_{Z\wt{f}_i \wt{f}_j} 
\tilde{f}_i^* \stackrel{\leftrightarrow}{\partial}_\mu \tilde{f}_j
\label{lag}
\end{eqnarray}
with the couplings
\begin{equation}
g_{Z\wt{f}_i \wt{f}_j}= -\frac{e}{s_Wc_W}  \{ 
(I_3^f -Q_f s_W^2) R_{j1}^{\wt{f}} R_{i1}^{\wt{f}} -   
Q_f s_W^2 R_{j2}^{\wt{f}} R_{i2}^{\wt{f}} 
\}  
\end{equation}
involving the matrix $R^{{\tilde{f}}}$ from the transformation~(\ref{eqe}).

\begin{figure}[t!]
\begin{center}
\vspace{-2.cm}
\input{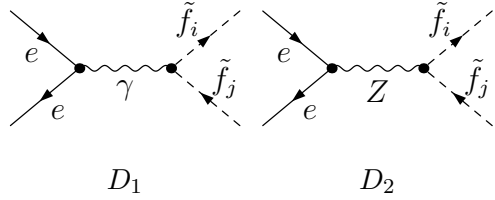}
\vspace{-13.7cm}
\caption{Tree level contribution to $e^+e^- \to \wt{f}_i \wt{f}_j^*$,
         with momenta $k_1(e^-)$, $k_2(e^+)$, $k_3(\wt{f}_i)$.
\label{born} }
\end{center}
\end{figure}

\vspace*{0.3cm}
Neglecting the electron--Higgs coupling, 
the production of sfermion pairs in $e^+ e^-$ 
collisions proceeds in lowest order through
$s$-channel photon and $Z$ boson exchanges (Fig.~\ref{born}). 
The tree level amplitude can be written as follows,
\begin{eqnarray}
{\cal M}^0 = {\cal M}_L^0 \ I_L + {\cal M}_R^0 \ I_R \quad {\mbox{with}} \quad 
I_{L,R}= \bar{v}(k_1)\not k_3 \frac{1\mp\gamma_5}{2} u(k_2) \label{inv}
\end{eqnarray}
and with
\begin{eqnarray}
{\cal M}_{L,R}^0 & = &  2 e^2 (-\delta_{ij} \frac{Q_f}{s} +
\frac{g_{L,R}\, g_{Z\wt{f}_i \wt{f}_j}}{(s-M_Z^2)} ) \, ,
\end{eqnarray}
where
\begin{eqnarray}
& & g_L= \frac{1-2 s_W^2}{2 s_W c_W} \qquad , 
\qquad g_R=\frac{-2 s_W^2}{2 s_W c_W} \, .
\end{eqnarray}
The angular distribution for the production of a
pair of sfermions $\wt{f}_i$, $\wt{f}_j^*$ is given by
\begin{eqnarray}
\frac{d\sigma_0}{d\Omega}=
N_C \frac{s \kappa_{ij}^3}{1024\pi^2 } ( |{\cal M}_L^0|^2 + |{\cal M}_R^0|^2) 
\sin^2\theta \, ,
\label{ang} 
\end{eqnarray}
where $N_C = 3$ (1) for scalar quarks (scalar leptons).
$\sqrt{s}$ is the CMS energy, $\theta$ the CMS scattering angle
between $e^+$ and $\wt{f}_i$, and
$\kappa_{ij}$ the final-sfermion velocity,
\begin{eqnarray}
\kappa_{ij}^2= (1-\mu_i^2 - \mu_j^2)^2 - 4 \mu_i^2 \mu_j^2
\ \ \ , \ \ \mu^2_{i,j}= \frac{m^2_{\wt{f}_{i,j}}}{s} \, .
\end{eqnarray}
Angular integration yields 
the total cross section,
\begin{eqnarray}
\sigma_0 =N_C\frac{s \kappa_{ij}^3 }{384 \pi} 
          (|{\cal M}_L^0|^2+|{\cal M}_R^0|^2) \, . 
\label{totalBorn}
\end{eqnarray}
The angular distribution~(\ref{ang}) is symmetric, and hence 
it is obvious that the forward--backward asymmetry 
$A_{FB}$, defined by
\begin{eqnarray}
A_{FB} = 
\frac{\int_{\theta \leq \pi/2} d\Omega \frac{d\sigma}{d\Omega}  
- \int_{\theta \geq \pi/2} d\Omega \frac{d\sigma}{d\Omega} } 
{\int_{\theta \leq \pi/2} d\Omega \frac{d\sigma}{d\Omega}  
+\int_{\theta \geq \pi/2} d\Omega \frac{d\sigma}{d\Omega} }=
\frac{\sigma_F-\sigma_B}{\sigma} \, ,
\end{eqnarray}
vanishes in lowest order. 
The quantum corrections, however, in particular the box diagrams, 
contribute to $A_{FB}\neq 0$.

\renewcommand{\theequation}{3.\arabic{equation}}
\setcounter{equation}{0}
\section{Structure of radiative corrections}

\subsection{One-loop diagrams for
$e^+e^- \to \wt{f}_i \wt{f}_j^*$}
\label{subsecdiags}

\begin{figure}[t!]
\begin{center}
\vspace{-2.cm}
\input{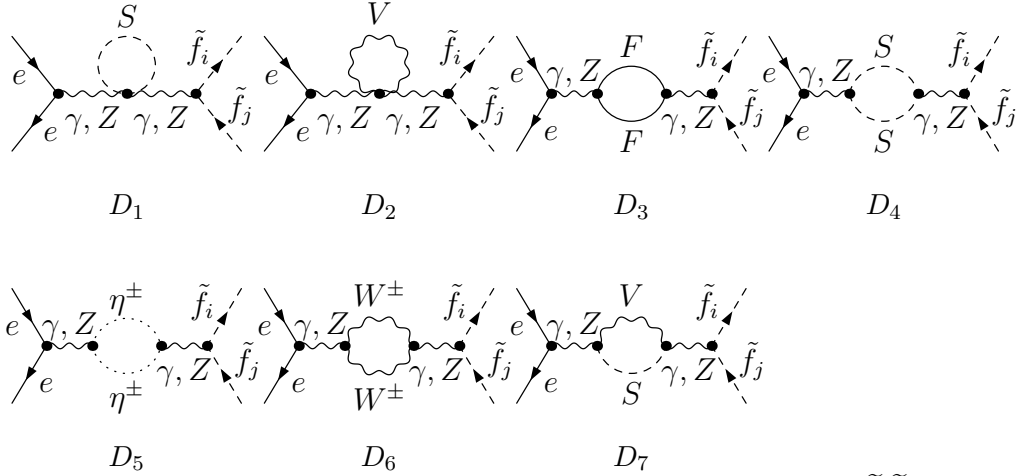}
\vspace{-10.5cm}
\caption{One--loop self--energy contributions to 
$e^+e^-\to \wt{f}_i \wt{f}_j^*$\label{self} }
\end{center}
\end{figure}

\begin{figure}[t!]
\begin{center}
\vspace{-2.6cm}
\input{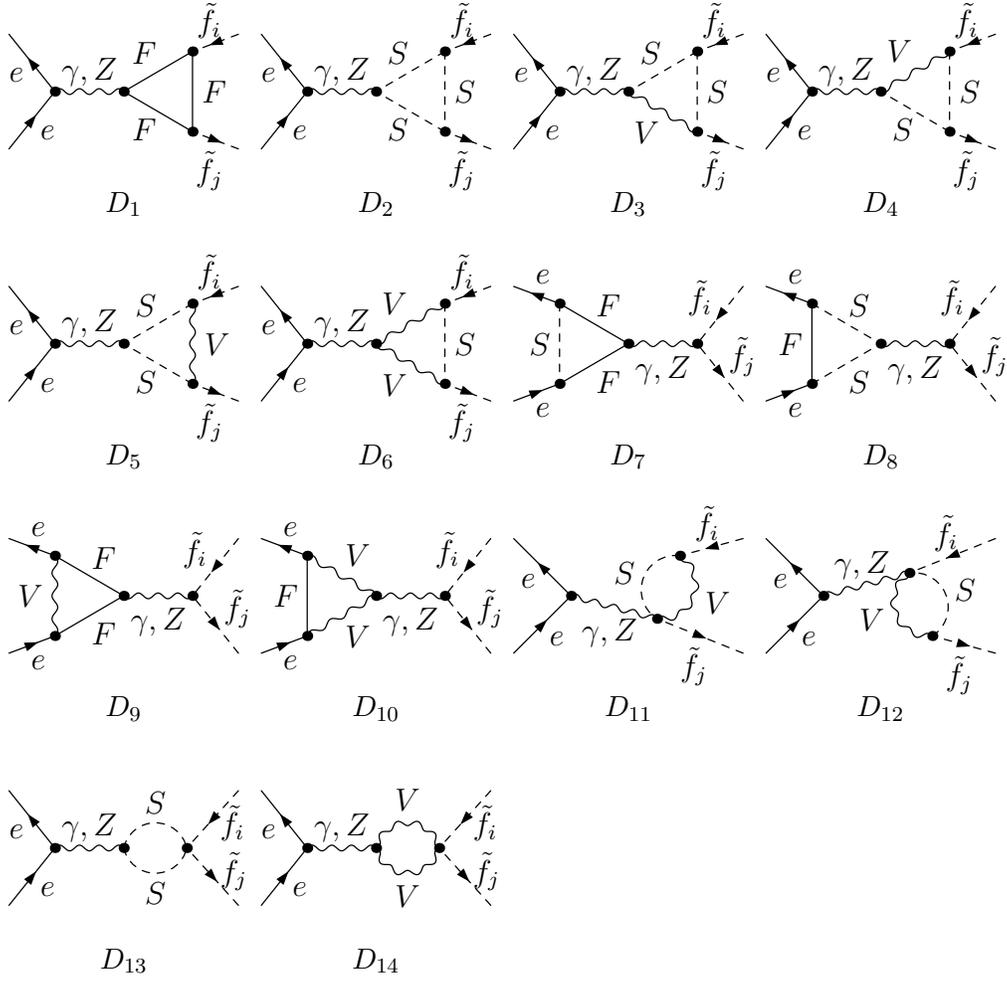}
\vspace{-3.3cm}
\caption{Vertex contributions to $e^+e^-\to \wt{f}_i \wt{f}_j^*$
\label{vertex} }
\end{center}
\end{figure}

\begin{figure}[b!]
\begin{center}
\vspace{-3.cm}
\input{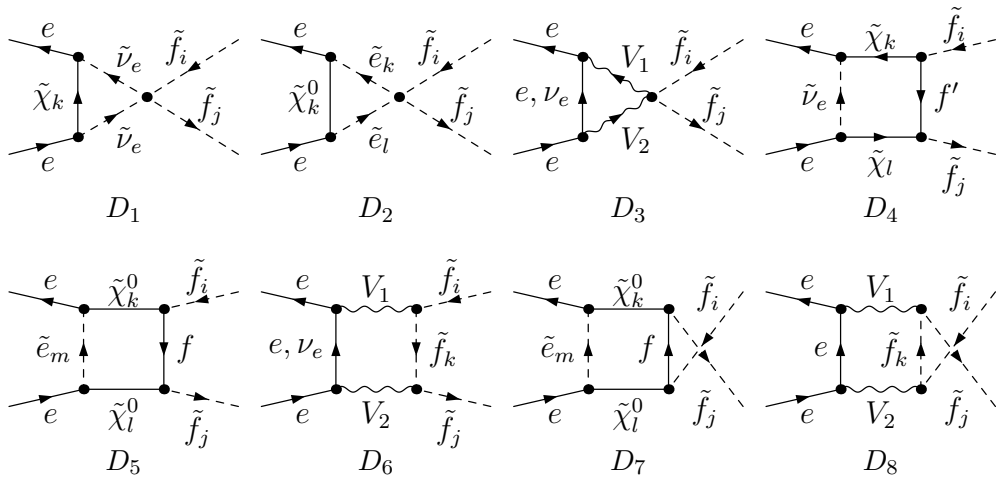}
\vspace{-10.cm}
\caption{Boxes contributions to $e^+e^-\to \wt{f}_i \wt{f}_j^*$\label{boxes} }
\end{center}
\end{figure}

The Feynman diagrams for the one-loop virtual contributions
are generically displayed in
Figures~\ref{self} -- \ref{counter}. 
These comprise the corrections to photon 
and $Z$ propagators and their mixing (Fig.~\ref{self}), the  corrections
to the initial-state vertices $\{\gamma,Z\} e^+e^-$, the 
corrections to the final-state vertices $\{\gamma, Z\} \wt{f}_i \wt{f}_j^*$ 
(Fig.~\ref{vertex}), 
and the box contributions (Fig.~\ref{boxes}).\footnote{In this terminology, 
diagrams with irreducible one-loop
4-point (3-point) functions are labeled as
box diagrams (vertex corrections)}
These diagrams are to be supplemented by the external 
self-energy contributions for 
$e^\pm$ and scalar fermions $\wt{f}_{i,j}$, which are part
of the counterterms for propagators and vertices 
(Fig.~\ref{counter}), to be added according to renormalization.
In the generic notation, $V,S,F$ denote all insertions of vector,
scalar, and fermionic states.

Note that loop contributions 
coming from initial state $e^+ e^- H_0$ and 
$e^+ e^- h_0$ vertices vanish for $m_e\rightarrow 0$
since $e^+$ and $e^-$ are both on-shell.
A similar argument holds for the
$A^0$ and neutral Goldstone-boson exchange diagrams.

The Feynman diagrams are generated and evaluated using the packages 
FeynArts and FormCalc~\cite{FA}. We have also used
LoopTools and FF~\cite{FF} in the numerical analysis.

The one-loop amplitudes are ultraviolet (UV) 
and infrared (IR) divergent. 
The UV singularities are treated by  dimensional
reduction \cite{siegel}
and are compensated
in the on-shell renormalization scheme. 
We have checked explicitly   
that the results are identical in using dimensional
reduction and dimensional regularization. 
The IR singularities are regularized with a 
small fictitious photon mass $\delta$. 


Diagrams like $D_{5,9}$ of Fig.~\ref{vertex} and diagrams $D_{6,8}$
of Fig.~\ref{boxes} and $D_{1\ldots 4}$ of Fig.~\ref{counter}
are IR divergent when the 
exchanged gauge boson is a photon. For an IR-finite
cross section we have to add  
the contribution from  real-photon emission,
$e^+ e^- \rightarrow 
\wt{f}_i \wt{f}_j^*\gamma$.
We limit ourself to the soft 
photon approximation where a cut $\Delta E$ on the energy of the photon 
is introduced. This approach have been used also for charged 
Higgs pair production in \cite{AM,GHA}.
  
It is interesting to note that the subclass of 
virtual and real photonic corrections
is UV finite and does not require renormalization of any parameters. 
UV divergences from virtual photons 
cancel between vertex corrections and external wave-function 
renormalizations. The photonic corrections are thus  
completely determined by the gauge couplings,
as already present in the Born approximation, 
and can be separately considered
as (cut-dependent) ``QED corrections''.

\subsection{Renormalization}
During the last decade, 
several approaches for renormalization of the MSSM 
in the on-shell scheme have been 
developed~\cite{gh}~--~\cite{itp} 
(see e.g.~\cite{vienna} for an overview).
Here we follow the strategy of~\cite{gh} by introducing 
counterterms for the physical parameters, i.e.\ for masses and
mixing angles, and perform  field renormalization
in a way that residues of renormalized propagators can be kept at unity.

For the  SM parameters and fields, 
we adopt the on-shell renormalization scheme
in the specification of~\cite{Denner}, with the following
replacements for the parameters 
\begin{eqnarray}
& & e \to (1+\delta Z_e) e \qquad , \qquad M^2_{W,Z} \to M^2_{W,Z} +
\delta M^2_{W,Z}\label{cd1} \, ,
\end{eqnarray}
and for the vector-boson fields
\begin{eqnarray}
Z \rightarrow Z_{ZZ}^{1/2} Z+
Z_{Z\gamma}^{1/2} A\, ,\quad  
A \rightarrow Z_{AA}^{1/2} A + Z_{\gamma Z}^{1/2} Z \, ,
\quad {\rm with} \quad
Z_{ab}^{1/2} = \delta_{ab} + \frac{1}{2} \delta Z_{ab}\, .
\label{cd2}
\end{eqnarray}
By the on-shell definition of the electroweak mixing angle
($s_W=\sin\theta_W$, $c_W=\cos\theta_W$) according to
$s_W^2=1-M_W^2/M_Z^2$, 
the corresponding counterterm is determined through the
$W$ and $Z$ mass  counterterms,
\begin{eqnarray}
\frac{\delta s_W^2}{s_W^2}=  \frac{c_W^2}{s_W^2} \Bigm ( 
\frac{\delta M_Z^2}{M_Z^2} - \frac{ \delta M_W^2}{M_W^2} \Bigm)\quad , \quad
\frac{\delta c_W^2}{c_W^2}=    
\frac{\delta M_Z^2}{M_Z^2} - \frac{\delta M_W^2}{M_W^2} \, .
\end{eqnarray}

\vspace*{0.3cm}
For the genuine SUSY part, we 
have to renormalize the scalar-fermion fields  
and the mixing angle $\wt{\theta}_{f}$ in (\ref{eqe}),
which is done by the substitution
\begin{eqnarray}
\widetilde{f}_1 \rightarrow Z_{11}^{1/2} \wt{f}_1+
Z_{12}^{1/2} \wt{f}_2 \ \ , \ 
\wt{f}_2 \rightarrow Z_{22}^{1/2} \wt{f}_2 + Z_{21}^{1/2} \wt{f}_1 \ \
, \ \
\heta \rightarrow \heta+\delta \heta 
\label{cd3}
\end{eqnarray}
with 
$ Z_{ij}^{1/2} = \delta_{ij} + \frac{1}{2} \delta Z_{ij}$.
\begin{figure}[t!]
\vspace{-2.cm}
\begin{center}
\input{counter.tex}
\vspace{-10.4cm}
\caption{Counterterms for renormalizing the 
         amplitudes for $e^+e^-\to \wt{f}_i \wt{f}_j^*$
\label{counter} }
\end{center}
\end{figure}

\vspace*{0.3cm}
Applying the prescriptions given above to the Lagrangian~(\ref{lag}) 
yields the counterterm Lagrangian for the SUSY vertices 
in Fig.~\ref{counter} as follows
(the other counterms in Fig.~\ref{counter} are formally 
the same as in the SM),
\begin{eqnarray}
 \delta {\cal L} &=&  
  A^\mu \sum_{i,j=1,2}\delta( \gamma \wt{f}_i\wt{f}_j) \; 
  \tilde{f}_i^* 
 \stackrel{\leftrightarrow}{\partial}_\mu
 \tilde{f}_j + 
 Z^\mu \sum_{i,j=1,2} \delta(Z \wt{f}_i\wt{f}_j) \;
\tilde{f}_i^* 
\stackrel{\leftrightarrow}{\partial}_\mu \tilde{f}_j
\label{dlag}
\end{eqnarray}
where
\begin{eqnarray}
 \delta( \gamma \wt{f}_i\wt{f}_j)&=& -e  Q_f (\frac{1}{2} \delta
  Z_{ij} +\frac{1}{2} \delta Z_{ji}) -e  Q_f (  \frac{1}{2}\delta
  Z_{\gamma\gamma} +  \delta Z_e) \delta_{ij}
+ \frac{1}{2} g_{Z\wt{f}_i \wt{f}_j} \  \delta Z_{Z\gamma} \, , 
\nonumber \\
\delta(Z \wt{f}_i\wt{f}_j) &=& -e Q_f \frac{1}{2} \delta Z_{\gamma Z} 
+ g_{Z\wt{f}_i \wt{f}_j} (\delta Z_e + \frac{1}{2} \delta Z_{ZZ} )
    \nonumber\\ & &
   + \frac{\delta s_W e }{c_W^3 s_W^2}
  [(-I_3^f - Q_f s_W^2 + 2 I_3^f s_W^2) R_{j1}^{\wt{f}} R_{i1}^{\wt{f}}
- Q_f s_W^2 R_{j2}^{\wt{f}} R_{i2}^{\wt{f}}  ]
\nonumber \\ & &
  +g_{Z\wt{f}_i \wt{f}_j}(\frac{1}{2} \delta Z_{ii} +
  \frac{1}{2} \delta Z_{jj}) +
  g_{Z\wt{f}_k \wt{f}_j}\delta Z_{ki} +g_{Z\wt{f}_i \wt{f}_l}
\delta Z_{lj} 
+ \Delta(g_{Z\wt{f}_i \wt{f}_j}) \delta \heta 
\end{eqnarray}
with $ \quad
\Delta(g_{Z\wt{f}_1 \wt{f}_1})=-\Delta(g_{Z\wt{f}_2 \wt{f}_2})=2
g_{Z\wt{f}_1 \wt{f}_2}\quad$ and $\quad \Delta(g_{Z\wt{f}_1
  \wt{f}_2})=\Delta(g_{Z\wt{f}_2 \wt{f}_1})= g_{Z\wt{f}_2
  \wt{f}_2}-g_{Z\wt{f}_1 \wt{f}_1}$.

\vspace*{0.5cm}
To fix all the renormalization constants,
the following conditions are imposed:

\begin{itemize}
\item The on-shell conditions for $M_W$, $M_Z$, $m_f$, $e$,
and for the gauge-field renormalization constants
as in the SM~\cite{Denner}.


\item On-shell conditions for the scalar fermions $\widetilde{f}_i$,
specified by the requirements of
mass renormalization, zero mixing on each mass shell, and residue =1 
for the diagonal sfermion propagators,
\begin{eqnarray}
\delta Z_{11} & = & \frac{\partial}{\partial p^2}
\Sigma_{\wt{f}_1 \wt{f}_1} (p^2) |_{p^2=m^2_{\wt{f}_1} }    \ \ , \
\   \delta Z_{22}= \frac{\partial}{\partial p^2}
\Sigma_{\wt{f}_2 \wt{f}_2} (p^2) |_{p^2=m^2_{\wt{f}_2} }   \ , \non \\
\delta Z_{12} & = &\frac{{\Sigma}_{\wt{f}_1 \wt{f}_2 }
(m_{\wt{f}_2}^2)}{m_{\wt{f}_2}^2- m_{\wt{f}_1}^2}\ ,   \  \delta Z_{21}=
\frac{{\Sigma}_{ \wt{f}_2 \wt{f}_1 }(m_{\wt{f}_1}^2)}
{m_{\wt{f}_1}^2-m_{\wt{f}_2}^2} \quad , \quad \delta
m^2_{\wt{f}_i}  =  \Sigma_{\wt{f}_i \wt{f}_j}
(m^2_{\widetilde{f}_i}) \, ,  \non \\
\delta\wt{\theta}_{{f}} & = & \frac{1}{2} \frac{\Sigma_{\wt{f}_1 \wt{f}_2}(
m^2_{ \wt{f}_2} )+ \Sigma_{\wt{f}_1 \wt{f}_2}(m^2_{ \wt{f}_1} ) }
{ m^2_{ \wt{f}_2} - m^2_{ \wt{f}_1} } \, ,
\label{renconditions}
\end{eqnarray}
where $\sum_{\wt{f}_i \wt{f}_j } (p^2)$, $i,j=1,2$ denotes the 
unrenormalized diagonal and non-diagonal sfermion self-energies.
The condition for the sfermion mixing angle has also been used 
in~\cite{djouadietal} in the context of QCD corrections to squark decays.
The sfermion-mass renormalization condition is listed for completeness;
it is not needed for our actual computation.

\end{itemize} 

\subsection{Amplitudes and cross sections}

In the limit of vanishing electron mass, 
the part of the amplitude following from the one-loop diagrams, 
${\cal M}^1$, can be  projected on the two invariants $I_{L,R}$
defined in eq.~(\ref{inv}),
\begin{equation}
 {\cal M}^1 = {\cal M}_L^1 \ I_L +  {\cal M}_R^1 \ I_R
   \ + \cdots 
\end{equation}
The omitted terms are of the type  
$\bar{v}(1\pm \gamma_5)u$; they vanish in the interference
with ${\cal M}_0$ and are hence not required at one-loop order.
For the counterterm part of the amplitude, $\delta {\cal M}^1 $, 
the projection on these two invariants is exact,
\begin{equation}
 \delta {\cal M}^1 =  \delta {\cal M}_L^1 \ I_L +  
\delta {\cal M}_R^1 \ I_R \, .
\end{equation}
The differential cross section can be written as follows,
\begin{equation}
  \frac{d \sigma_1}{d \Omega} =
  \frac{d \sigma_0}{d \Omega} +
\{\ 2\, \Re [ {\cal M}_0^{*}( {\cal M}^1+  \delta {\cal M}^1) ] + 
|{\cal M}^1+  \delta {\cal M}^1|^2\ \} \cdot
\frac{1}{4} \,\frac{\kappa_{ij}}{64 \pi^2 s} 
\label{dif}
\end{equation}
(spin summation to be understood)
with the Born cross section from~(\ref{ang}).
At one-loop order, only the interference term contributes,
\begin{equation}
 \Re [ {\cal M}_0^{*} ({\cal M}^1 +  \delta {\cal M}^1) ] = 
[ {\cal M}_L^{0} ( {\cal M}^1_L+  \delta {\cal M}^1_L) +
  {\cal M}_R^{0} ( {\cal M}^1_R+  \delta {\cal M}^1_R) 
] \,
\frac{s^2}{4} \kappa_{ij}^2 \sin^2\theta \, ,
\label{square1}
\end{equation}
on which our numerical analysis is based.
Nevertheless, we will comment also on the purely-quadratic term,
\begin{eqnarray}
& & |{\cal M}^1+  \delta {\cal M}^1|^2 =(|{\cal M}_L^{1}+\delta {\cal
  M}_L^{1}|^2  + |{\cal M}_R^{1}+\delta {\cal M}_R^{1}|^2) \,
\frac{s^2}{4} \kappa_{ij}^2 \sin^2 \theta \, ,
\label{square2}
\end{eqnarray}
which may be useful to give some partial 
information on the size of 
higher-order corrections of ${\cal O}(\alpha^2)$.

\vspace*{0.2cm}
The integrated  cross section at the one-loop level, $\sigma_1$,
derived from (\ref{dif}) with the interference term only, 
can be written in the following way,
\begin{eqnarray}
\sigma_1=\sigma_0 + \sigma_0\Delta \, ,
\label{sig1}
\end{eqnarray}
pointing out the relative correction 
\begin{eqnarray}
\Delta = (\sigma_1-\sigma_0 )/\sigma_0
\label{relat}
\end{eqnarray}
with respect to the Born cross section
$\sigma_0$ in eq.~(\ref{totalBorn}).

The relative correction $\Delta$ can be 
decomposed  into the following parts, indicating their origin,
\begin{eqnarray}
\Delta =\Delta_{\mbox{self}}+\Delta_{\mbox{vertex}}+\Delta_{\mbox{boxes}}+
\Delta_{\mbox{QED}}+\Delta_{\mbox{SUSY-QCD}}\label{split}
\end{eqnarray}
where $\Delta_{\rm SUSY-QCD}$ denotes the supersymmetric QCD corrections 
with virtual gluinos as well as 
virtual and real gluons. The gluino part has been 
recalculated here in connection
with the electroweak contributions, and the gluon part
is taken  over from~\cite{ACD}.

The electroweak terms $\Delta_{\rm vertex}$ and $\Delta_{\rm boxes}$ do
not contain any virtual-photon diagram. 
According to the discussion at the end of section~\ref{subsecdiags},
all virtual-photon diagrams 
for vertex, box, and external wave-function contributions 
have been properly separated and combined with real bremsstrahlung
to form the subclass of QED corrections,
described by $\Delta_{\rm QED}$. 

We note that universal-type quantities, like self energies,
initial and final state vertex corrections, do not generate any
forward--backward asymmetry.  
Indeed, $A_{FB}$ is only generated by non the universal part,
essentially from box diagrams and real bremsstrahlung, as discussed later.

For the loop-induced forward--backward asymmetry $A_{FB}$ we will
apply the same definition as in \cite{GHA},
\begin{eqnarray}
A_{FB}=\frac{\sigma_F-\sigma_B}{\sigma}=
\frac{\sigma_F-\sigma_B}{\sigma_0} \frac{1}{1+\Delta}
=\Delta_{FB}\frac{1}{1+\Delta} \, .
\label{FBA}
\end{eqnarray}
$\Delta_{FB}$ is the antisymmetric part of the cross section
normalized to the Born cross section~$\sigma_0$.


\renewcommand{\theequation}{4.\arabic{equation}}
\setcounter{equation}{0}
\section{Numerical evaluation and discussion}

To set the basis for the numerical evaluation,
we specify the free parameters that will be used. 
\begin{itemize}
\item The MSSM Higgs sector is 
parametrized by the CP-odd mass $M_A$ and $\tan\beta$, taking into account 
radiative corrections with the help of
FeynHiggs~\cite{HHS}, and we assume  
$\tan\beta \geq 2.5$.
 
\item The chargino--neutralino sector can be parametrized by the 
gaugino-mass  terms
$M_1$, $M_2$, and the Higgsino-mass term $\mu$. For simplification  
we assume $M_1\approx M_2/2$. 

\item
Sfermions are characterized 
by a common soft-breaking sfermion mass 
$M_{SUSY} \equiv \widetilde{M}_L=\widetilde{M}_R$ and
soft trilinear couplings $A_f$. Again, for simplicity,  
we will take them as uniform 
for $\tilde{t}$, $\tilde{b}$,  and $\tau$
($A_0 \equiv A_t=A_b=A_\tau$).
\end{itemize}
As experimental data points~\cite{pdg},
the following input quantities enter:
$\alpha^{-1}=137.03598$, $m_Z=91.1875$ GeV, 
$m_W=80.45$ GeV, $m_t = 174.3$ GeV.
 

\begin{figure}[t!]
\smallskip\smallskip 
\vskip-4.4cm
\centerline{{
\epsfxsize2.99 in 
\epsffile{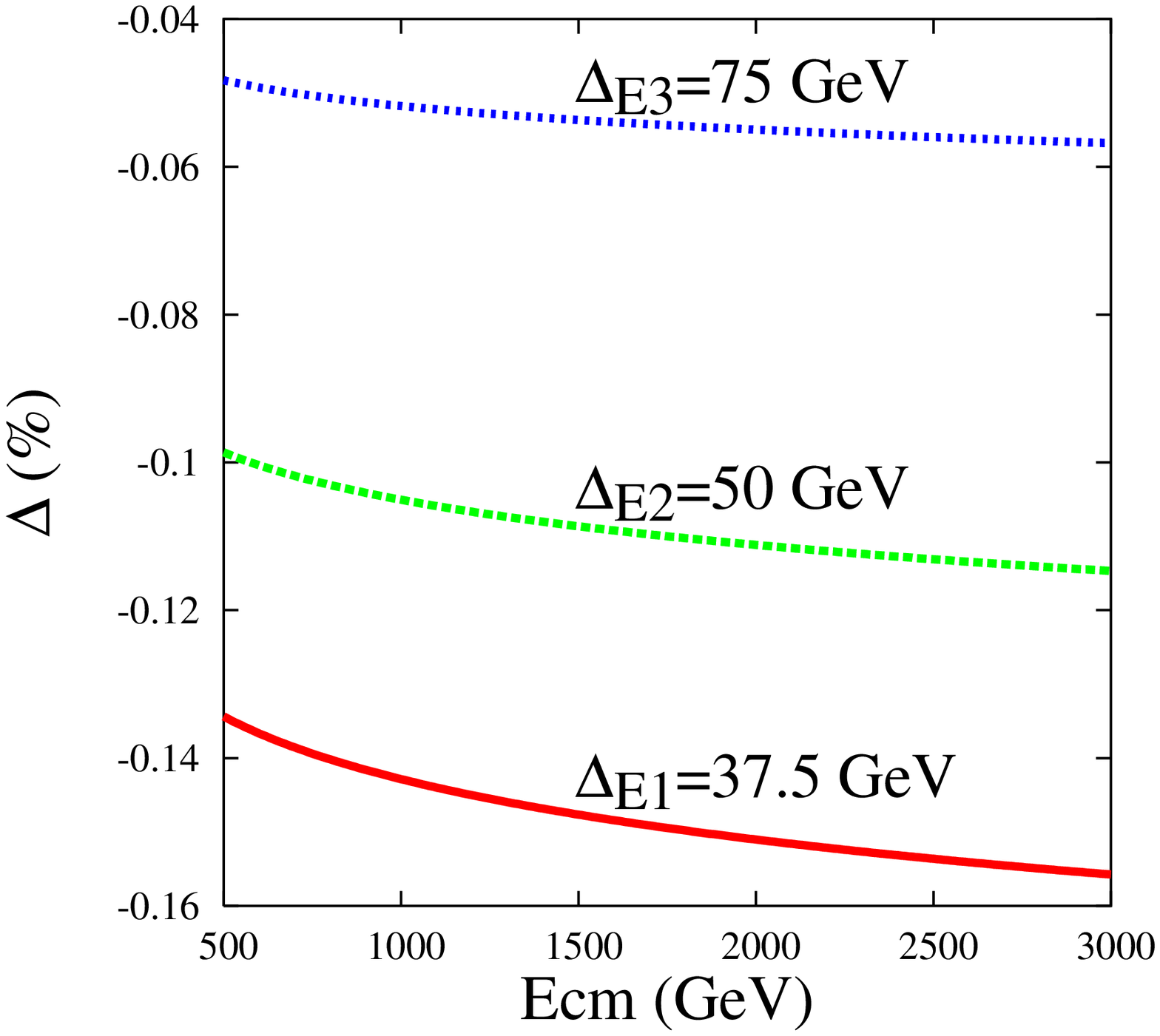}}  \hskip0.4cm
\epsfxsize3.05 in 
\epsffile{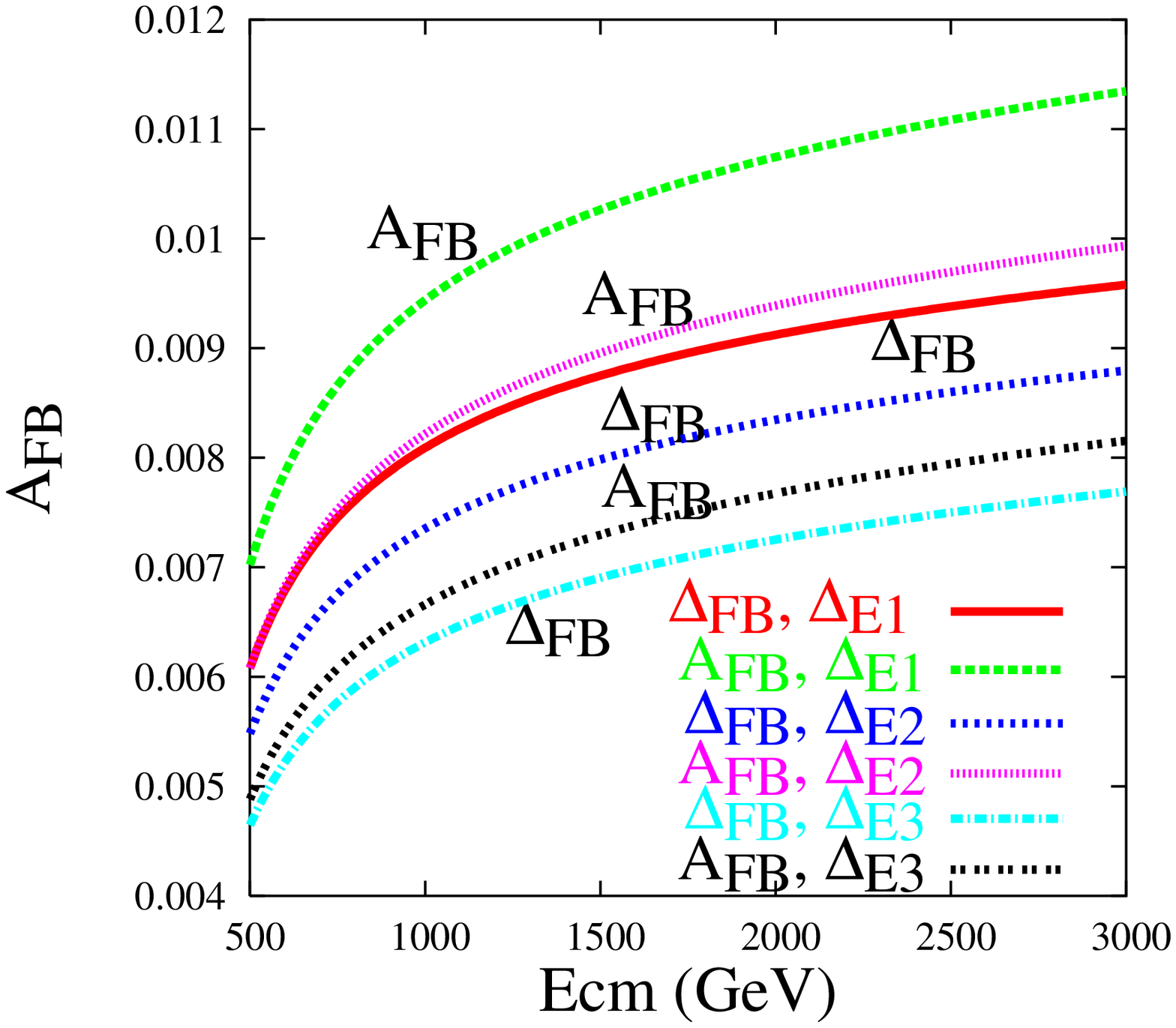} }
\caption{QED corrections to the total cross section (left) 
and the forward-backward asymmetry (right)
for $e^+e^- \to \wt{b}_1 \wt{b}_1^*$ in
scenario $sc_1$, as function of the CM energy
$E_{\rm CM} = \sqrt{s}$. }
\label{qed}
\end{figure}

Before presenting our results, we 
want to mention that we  performed cross checks with the 
results of~\cite{helmut2} for the subclass of Yukawa corrections
considered there and found perfect agreement for $\tilde{t}$ 
and $\tilde{b}$ production
\footnote{ For $\tilde{b}$
production we got agreement 
after correcting an overall sign in the charged Higgs 
contribution in~\cite{helmut2}.}.
Furthermore, QED corrections, 
self energies, boxes, and SUSY-QCD corrections have been 
checked with independent calculations not based on FormCalc. 

\vspace*{0.2cm}
For illustration of the effects of the radiative corrections, 
we have chosen the following three  scenarios:
\begin{itemize}
\item[$sc_1$:] $\tan\beta=6$, $M_A=250$, $M_{SUSY}=200$, $\mu=800$,
$M_2=200$ and $A_0=400$ GeV.
$m_{\wt{\tau}_{1,2}}=(185,223)$ GeV,
$m_{\wt{t}_{1,2}}=(148,339)$ GeV,
$m_{\wt{b}_{1,2}}=(146, 250)$ GeV, \\
$m_{\wt{c}_{1,2}}=m_{\wt{c}_{L,R}}=(193,197)$ GeV,
$m_{\wt{s}_{1,2}}=m_{\wt{s}_{R,L}}=(201,209)$ GeV;
$m_{\tilde{g}} = 525$ GeV.
\item[$sc_2$:] $\tan\beta=18$, $M_A=250$, 
 $M_{SUSY}=400$, $\mu=1600$,
 $M_2=200$ and $A_0=800$ GeV.
$m_{\wt{\tau}_{1,2}}=(335,460)$ GeV,
$m_{\wt{t}_{1,2}}=(254,559)$ GeV,
$m_{\wt{b}_{1,2}}=(175,542)$ GeV, \\
$m_{\wt{c}_{1,2}}=m_{\wt{c}_{L,R}}=(396,399)$ GeV,
$m_{\wt{s}_{1,2}}=m_{\wt{s}_{R,L}}=(397,408)$ GeV;
$m_{\tilde{g}} =524$ GeV.
\item[$sc_3$:] 
$\tan\beta=30$, $M_A=250$, $M_{SUSY}=200$, $\mu=200$, 
$M_2=1000$ and $A_0=300$ GeV. $m_{\wt{\tau}_{1,2}}=(179,228)$ GeV,
$m_{\wt{t}_{1,2}}=(131,346)$ GeV,
$m_{\wt{b}_{1,2}}=(124,263)$ GeV, \\
$m_{\wt{c}_{1,2}}=m_{\wt{c}_{L,R}}=(192,197)$ GeV,
$m_{\wt{s}_{1,2}}=m_{\wt{s}_{R,L}}=(201,209)$ GeV;
$m_{\tilde{g}} = 2624$ GeV.
\end{itemize}
$sc_1$ is gaugino-like with rather light sfermions,
and $sc_2$ is also gaugino-like with intermediate $\tan\beta$,
and heavier sfermions, and
$sc_3$ is higgsino-like with large $\tan\beta=30$.

As explained before, QED corrections can be
isolated and can be studied separately.
The corrections depend on the values of the photon-energy cut
$\Delta E$. More realistic situations require more complicated
cuts and a more sophisticated treatment of the real photon part.
Nevertheless, we display $\Delta_{\rm QED}$ in  Fig.~\ref{qed}
to indicate the size of the QED effects, for various energy cuts.
Fig.~\ref{qed} is for scenario $sc_1$ and for the special case of 
$\tilde{b}_1$-pair production. 
The behavior for other scenarios and fermion states
is almost the same, as long as one is away from the thresholds
localized at different energies,
where the Coulomb singularity shows up. 
Also given in Fig.~\ref{qed}  
is the QED-induced forward-backward asymmetry,
which is very small, well below 1\% up to the TeV range.
In the case of QED corrections,
large double-logarithms of the Sudakov type are absent due  
to  cancellations by the bremsstrahlung contributions.

Our main emphasis will be
on the residual set of non-QED corrections, which involve the
quantum structure of the complete MSSM. 
In all the figures  of the following discussions, 
the QED corrections are not included. \\

\noindent
{\bf Third-generation squarks}

\begin{figure}[t!]
\smallskip\smallskip 
\vskip-4.3cm
\centerline{{
\epsfxsize2.8 in 
\epsffile{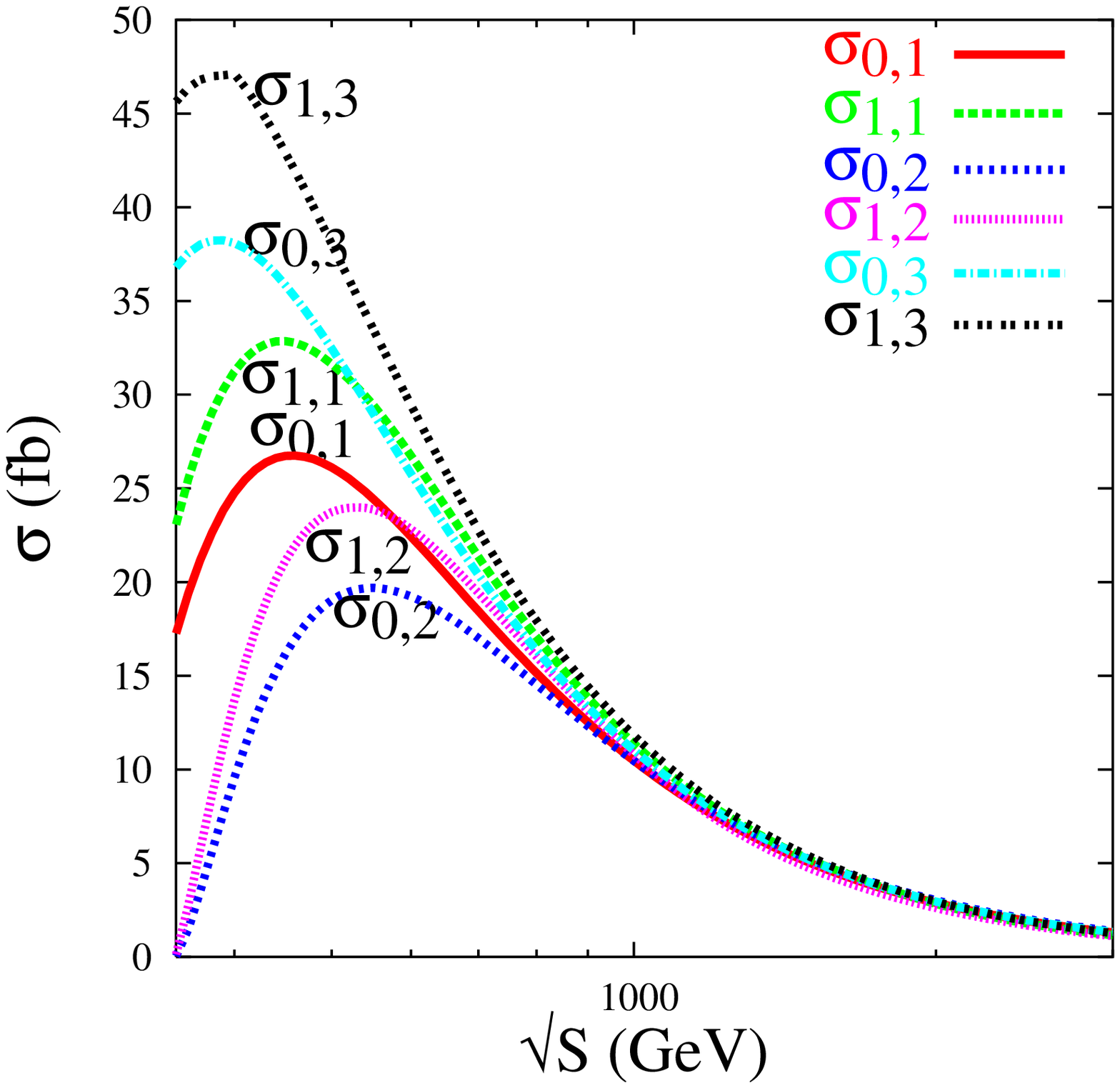}}  \hskip0.4cm
\epsfxsize2.8 in 
\epsffile{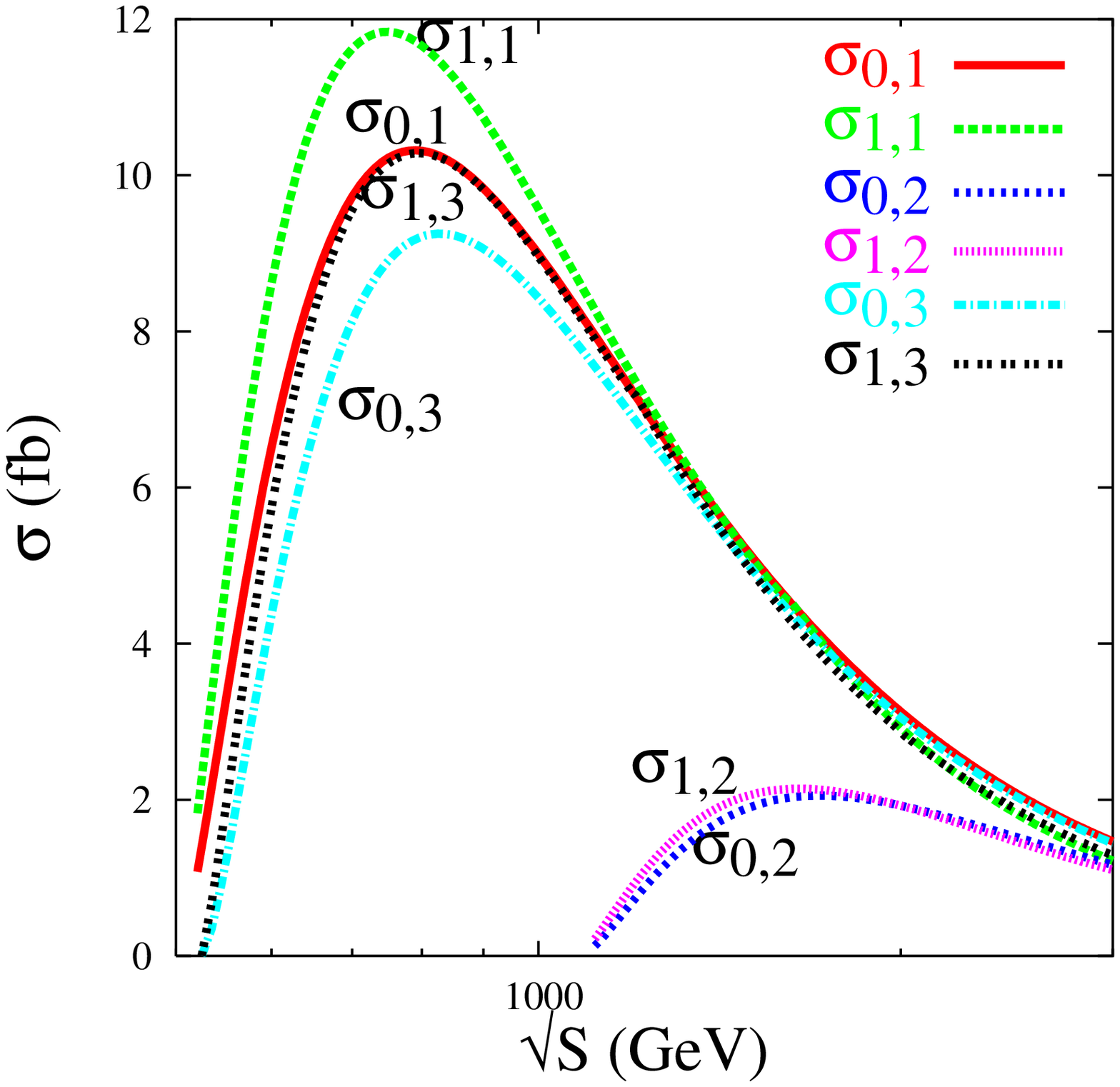} }
\vskip-3.5cm
\centerline{{
\epsfxsize2.9 in 
\epsffile{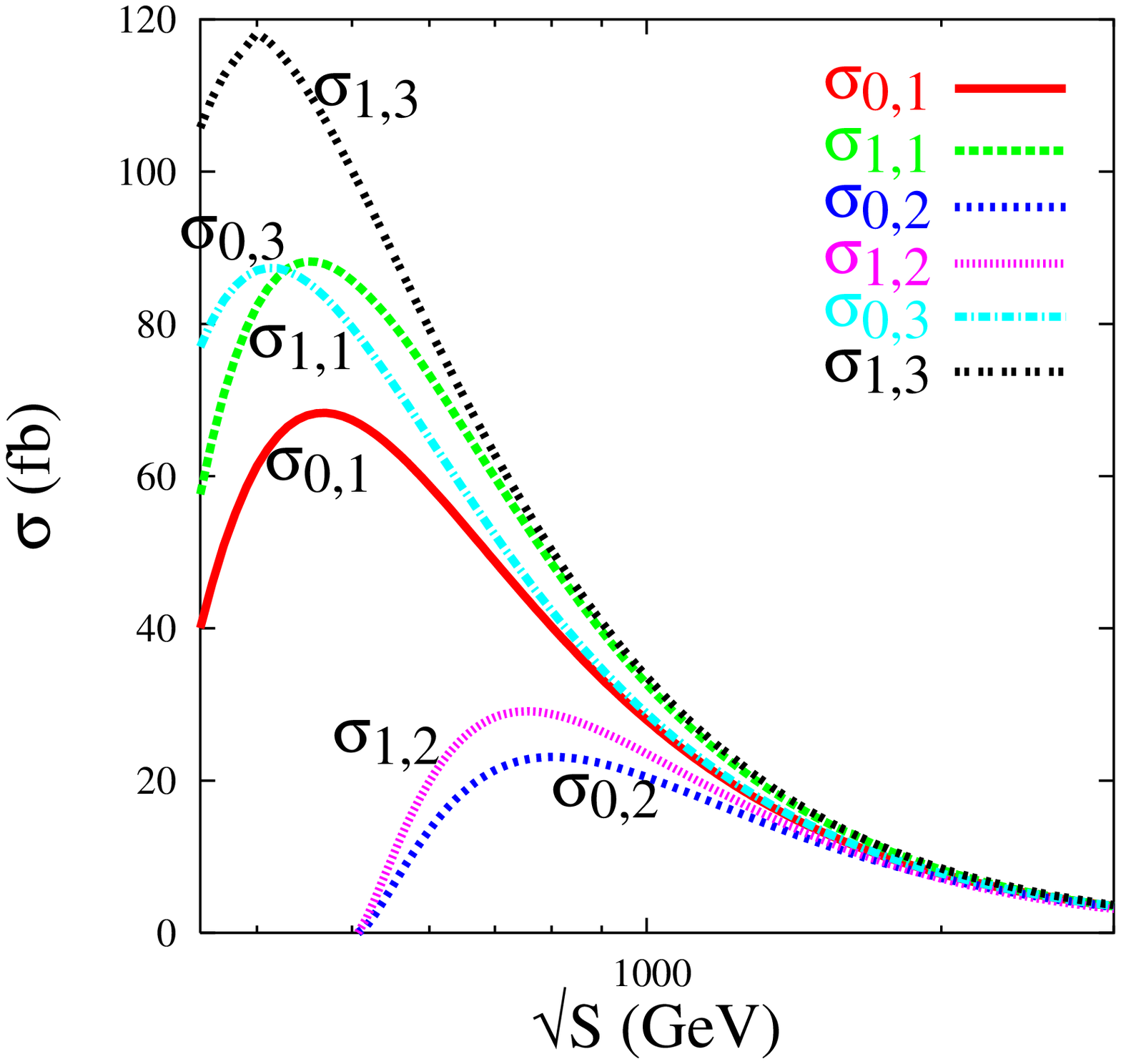}}  \hskip0.4cm
\epsfxsize2.8 in 
\epsffile{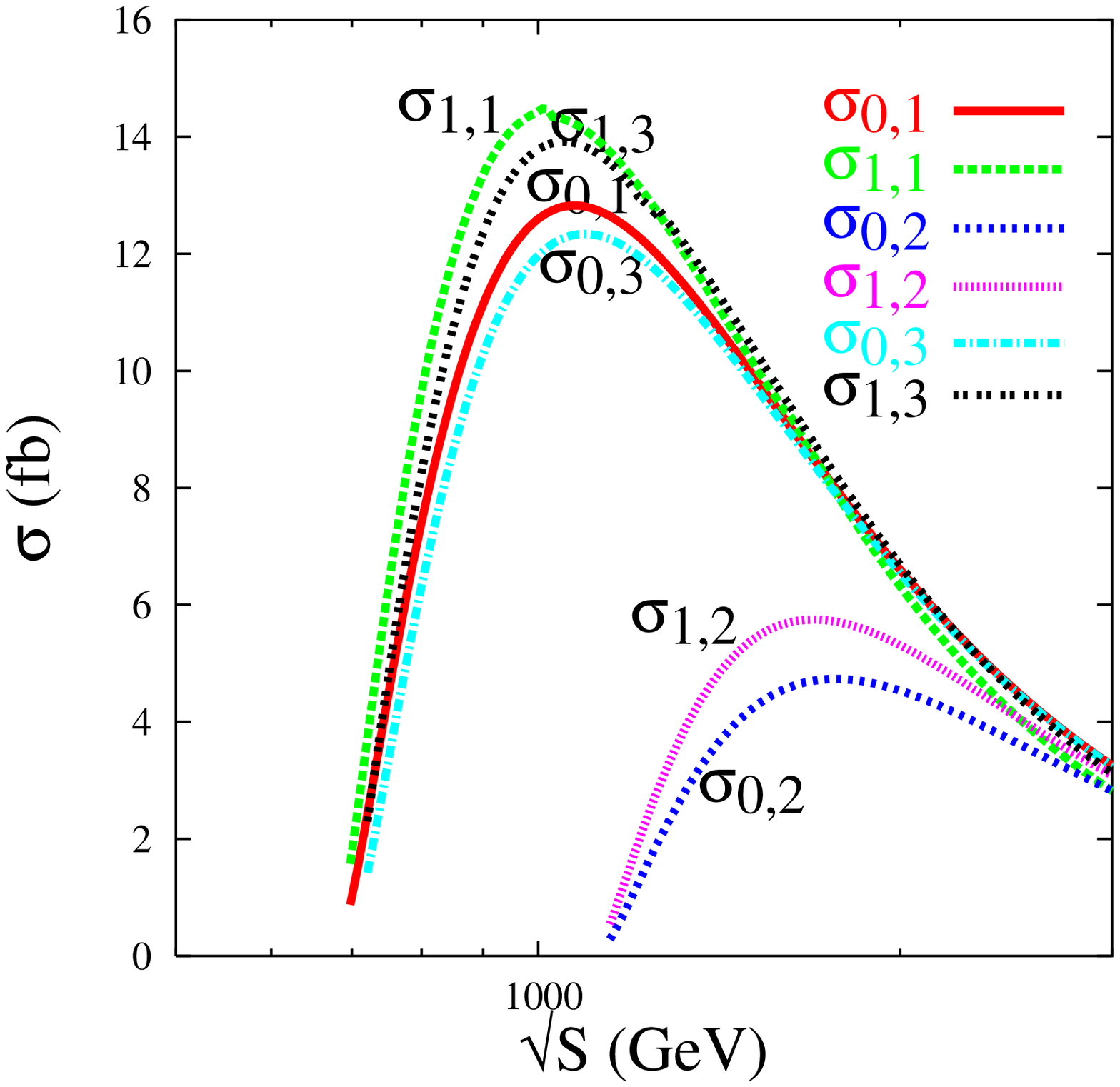} }
\caption{Integrated tree-level cross sections
$\sigma_{0,i}$ and 
one-loop cross section $\sigma_{1,i}$ in the three scenarios
($i=sc_1,sc_2,sc_3$) for $e^+e^- \to \wt{b}_{k}\wt{b}_{k}^*$ 
(upper part) and $e^+e^- \to \wt{t}_{k}\wt{t}_{k}^*$ (lower part).
The lighter squarks ($k=1$) are on the left, the heavier squarks
($k=2$) on the right side. }
\label{csbt}
\end{figure}
In Fig.~\ref{csbt} we show 
the integrated tree-level cross section $\sigma_{0,i}$ 
for our three scenarios ($i=sc_1,sc_2,sc_3$) and the corresponding 
one-loop cross section $\sigma_{1,i}$ for
the squarks of the third generation ($\tilde{q}=\tilde{b}, \tilde{t}$),
$e^+e^- \to \wt{q}_1\wt{q}_1^*$ and
$e^+e^- \to \wt{q}_2\wt{q}_2^*$, 
as function of the center of mass energy.
As it can be seen in Fig.~\ref{csbt}, 
in the energy range where the cross sections are large,
the loop contributions lead to a further significant 
enhancement.

\begin{figure}[t!]
\smallskip\smallskip 
\vskip-3.4cm
\centerline{{
\epsfxsize2.765 in 
\epsffile{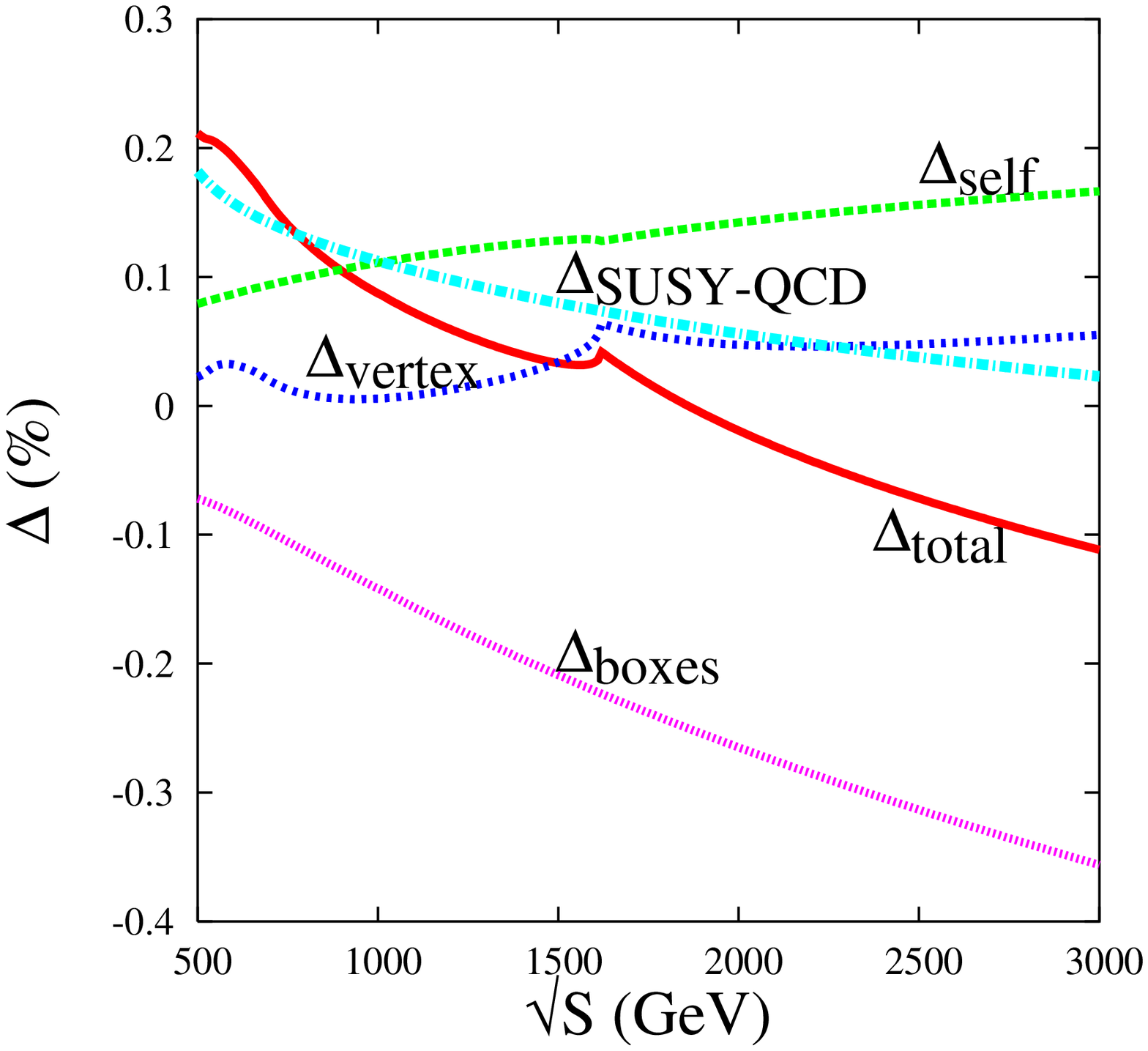}}  \hskip0.4cm
\epsfxsize2.765 in 
\epsffile{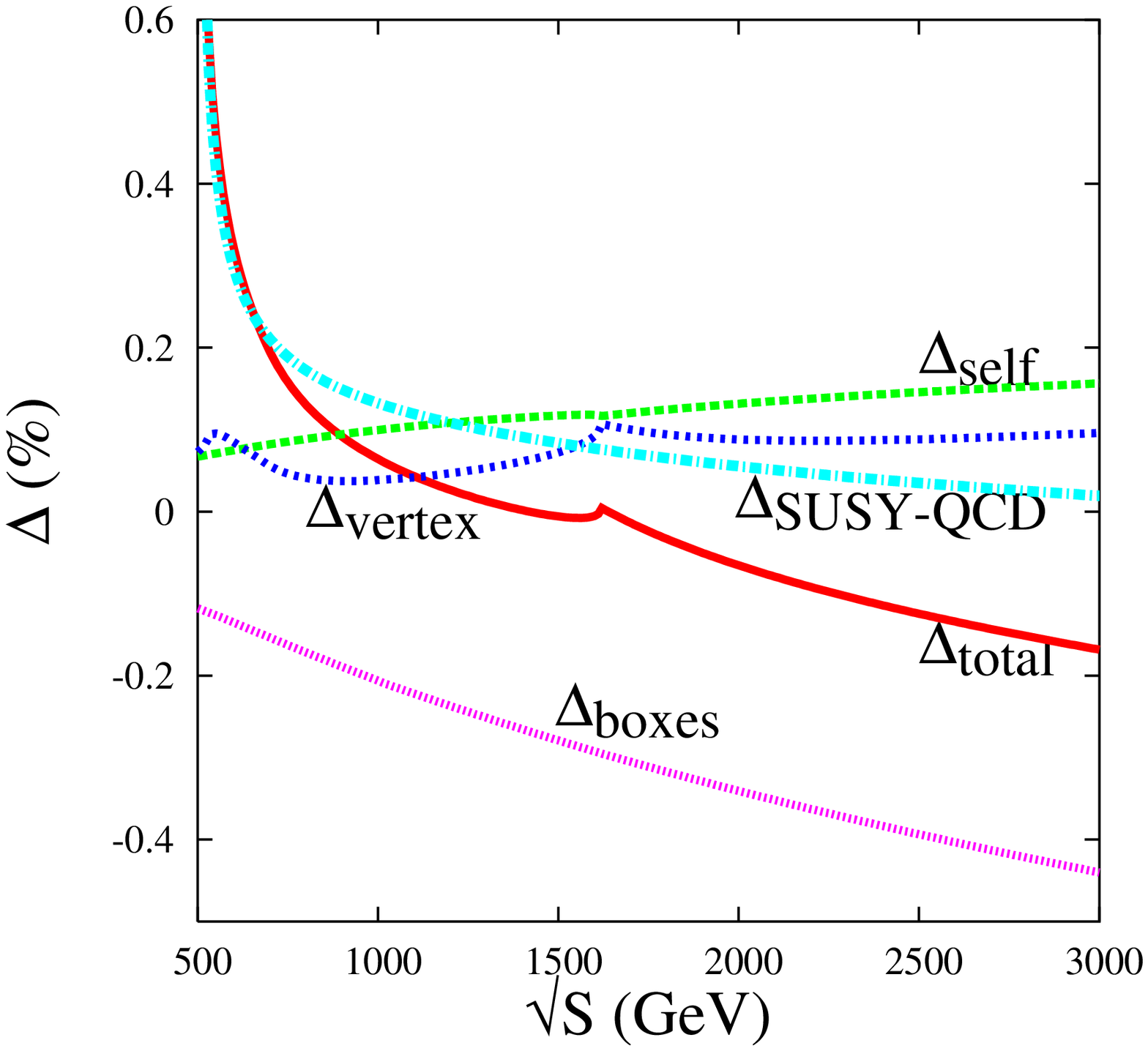} }
\vskip-3.6cm
\centerline{{
\epsfxsize2.765 in 
\epsffile{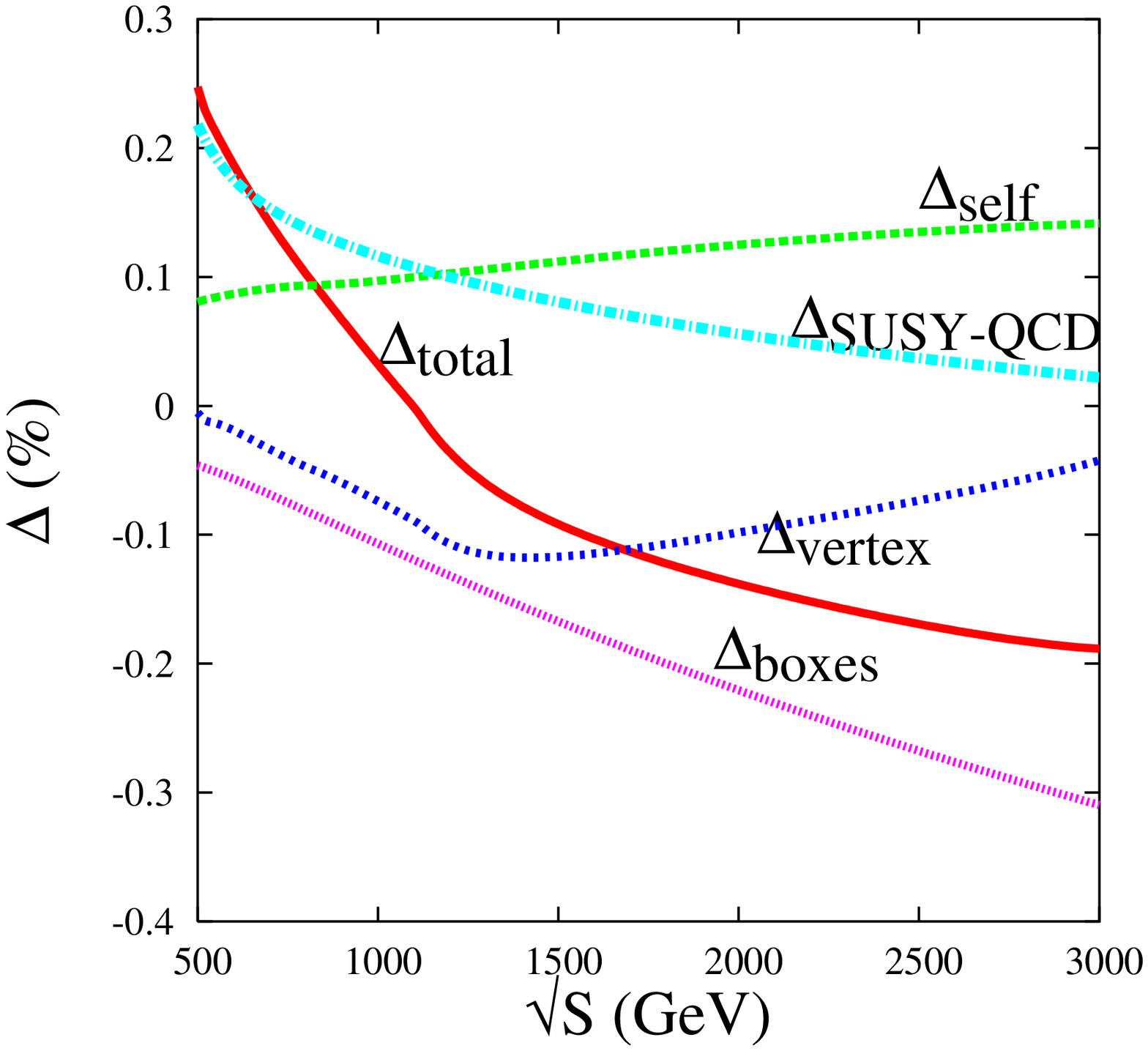}}  \hskip0.4cm
\epsfxsize2.765 in 
\epsffile{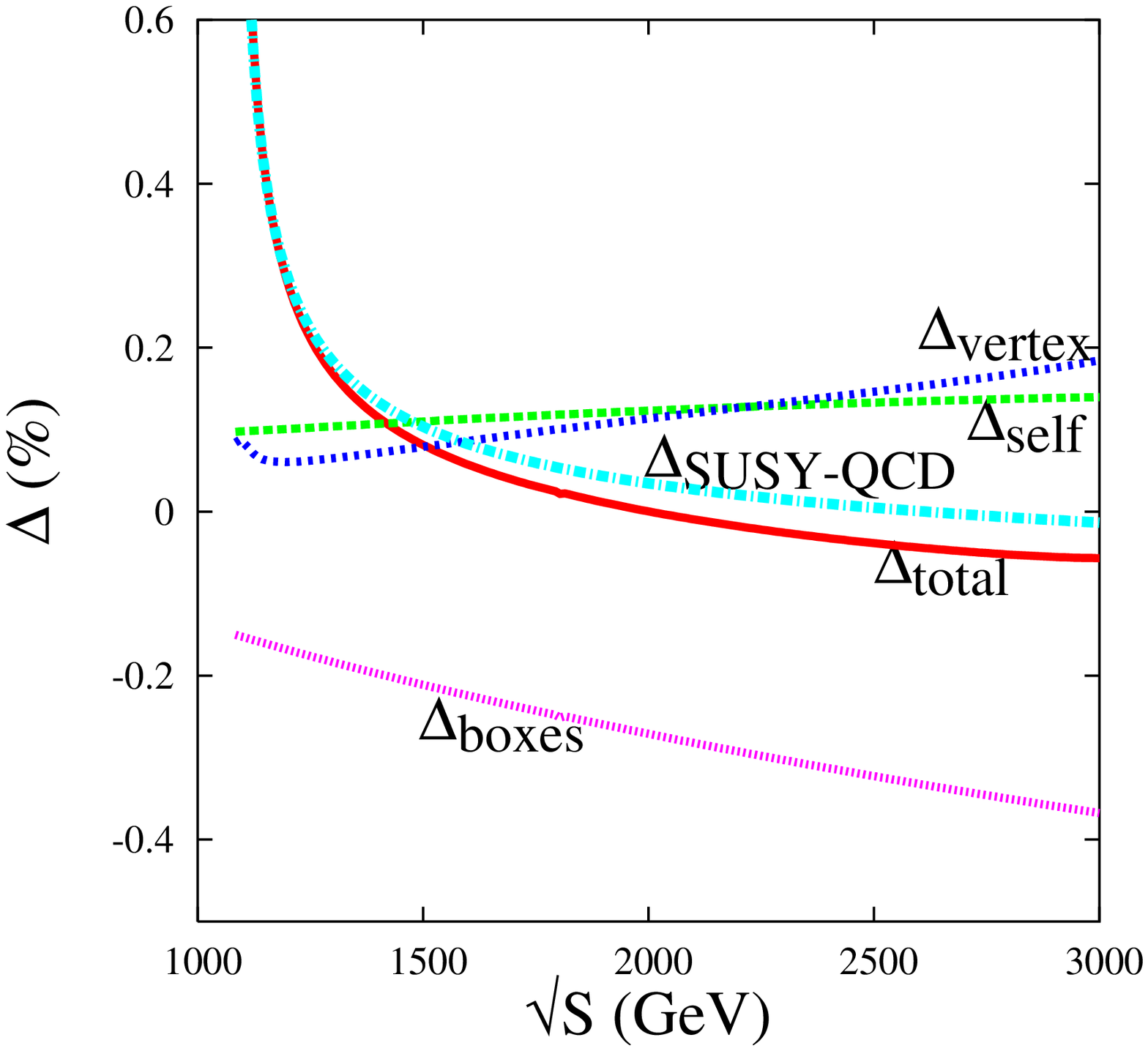} }
\vskip-4.cm
\smallskip\smallskip 
\centerline{{
\epsfxsize2.765 in 
\epsffile{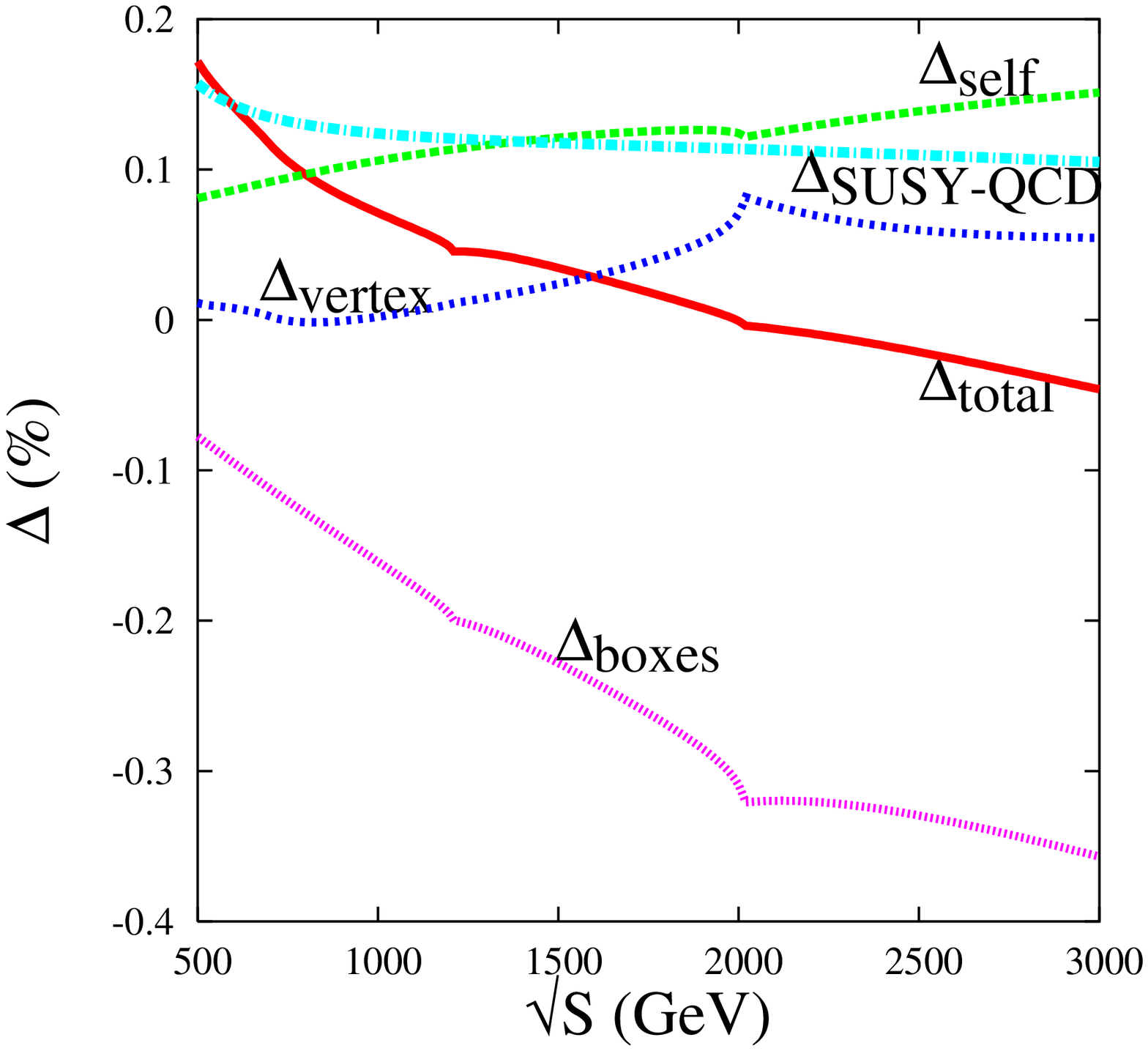}}  \hskip0.4cm
\epsfxsize2.765 in 
\epsffile{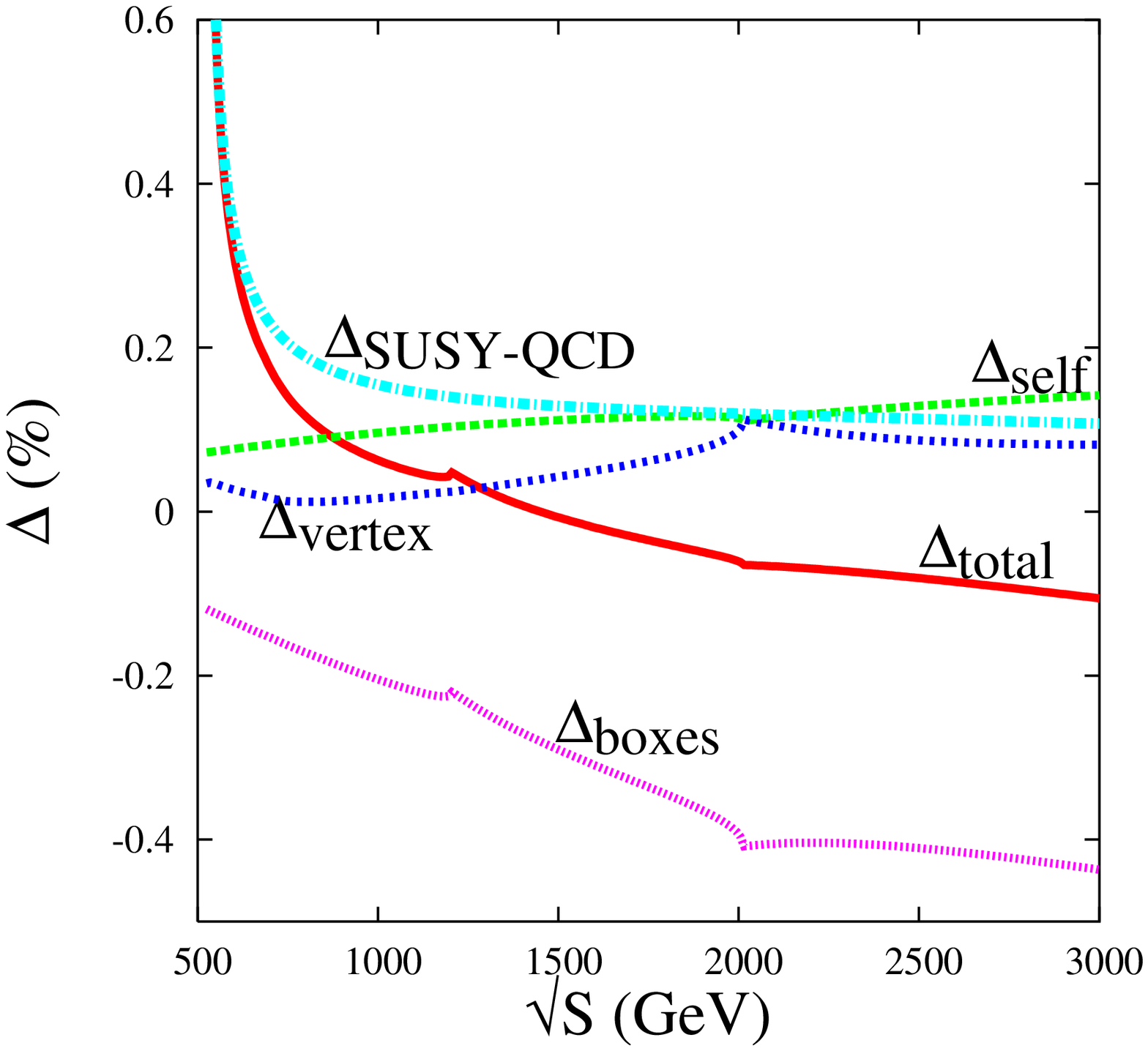} }
\smallskip\smallskip
\caption{Electroweak and SUSY-QCD corrections to 
$e^+e^- \to \wt{b}_1\wt{b}_1^*$ (left) and $e^+e^- \to \wt{b}_2\wt{b}_2^*$ 
(right). Scenario $sc_1$ (upper plots), $sc_{2}$ (middle plots), 
and $sc_{3}$ (lower plots).}
\label{del1}
\end{figure}

\begin{figure}[t!]
\smallskip\smallskip 
\vskip-3.4cm
\centerline{{
\epsfxsize2.765 in 
\epsffile{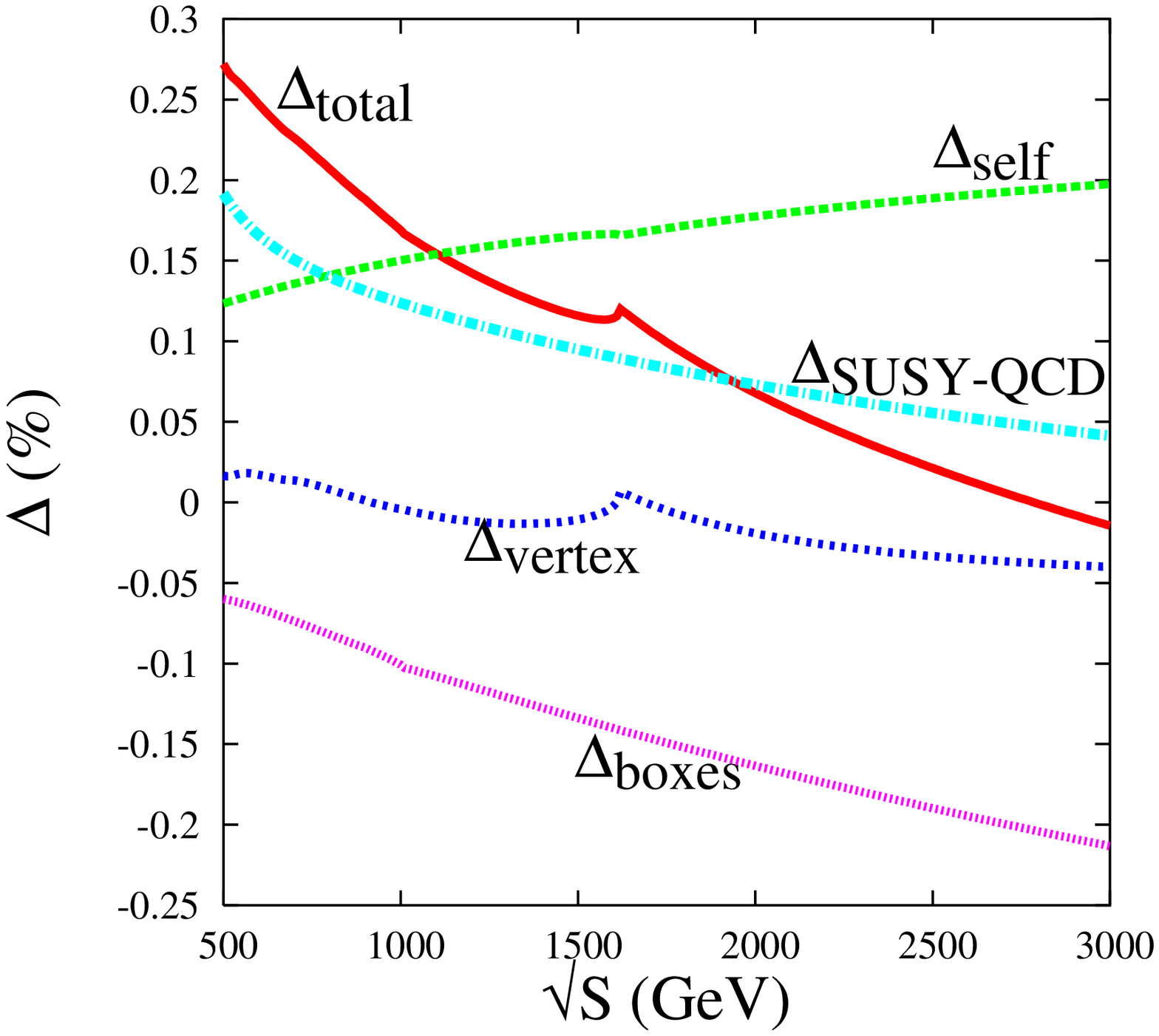}}  \hskip0.4cm
\epsfxsize2.765 in 
\epsffile{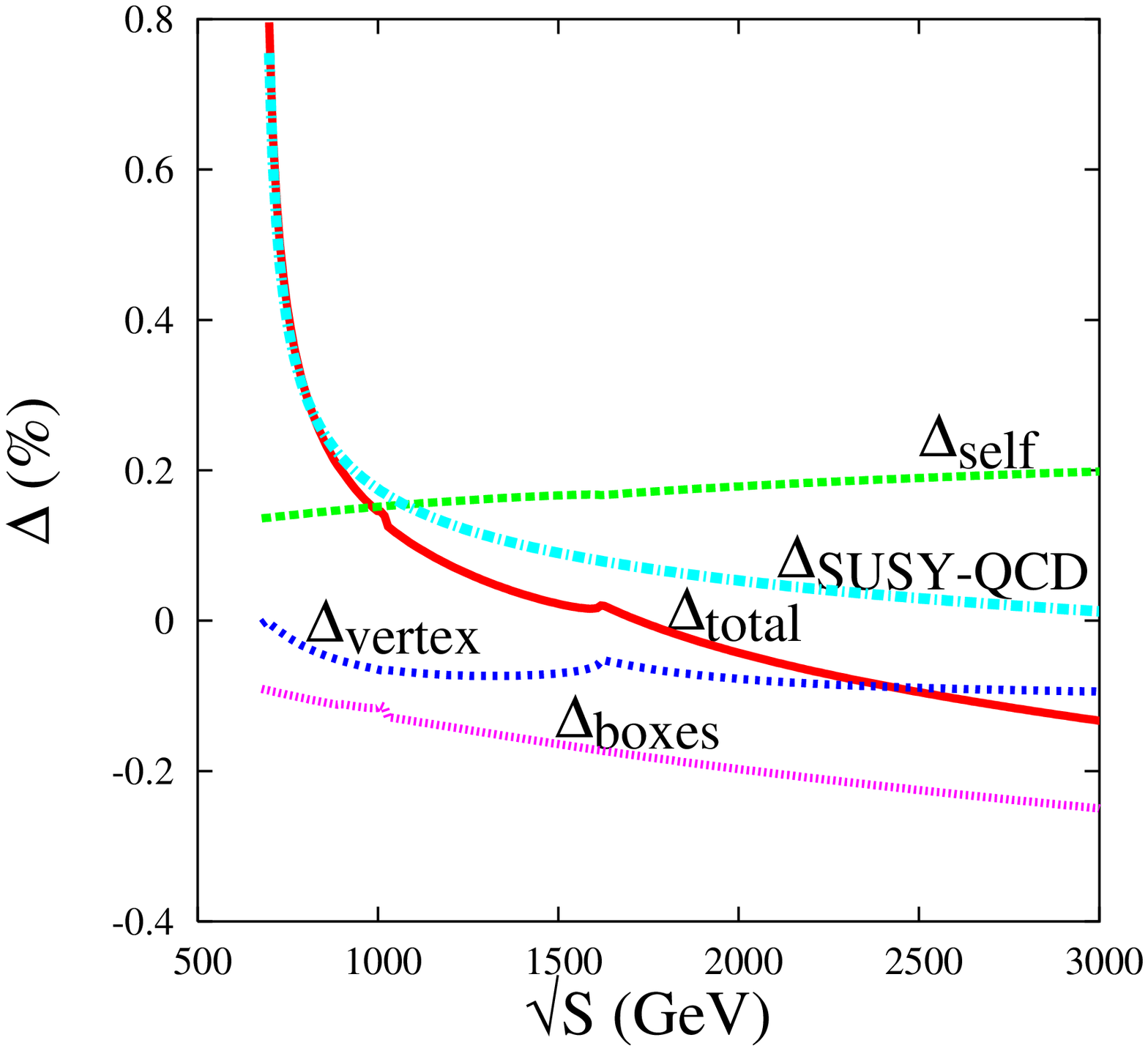} }
\vskip-3.6cm
\centerline{{
\epsfxsize2.765 in 
\epsffile{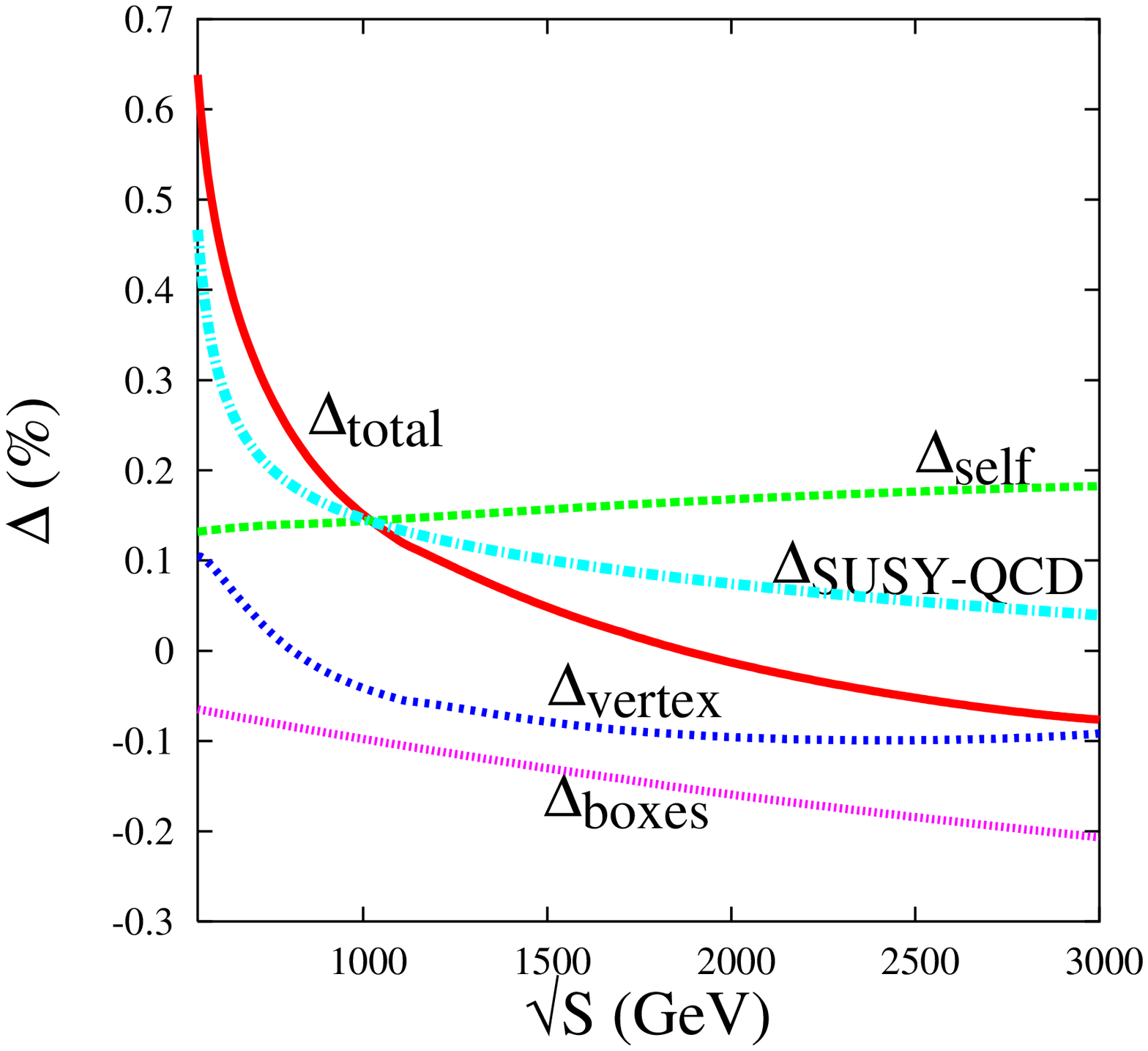}}  \hskip0.4cm
\epsfxsize2.765 in 
\epsffile{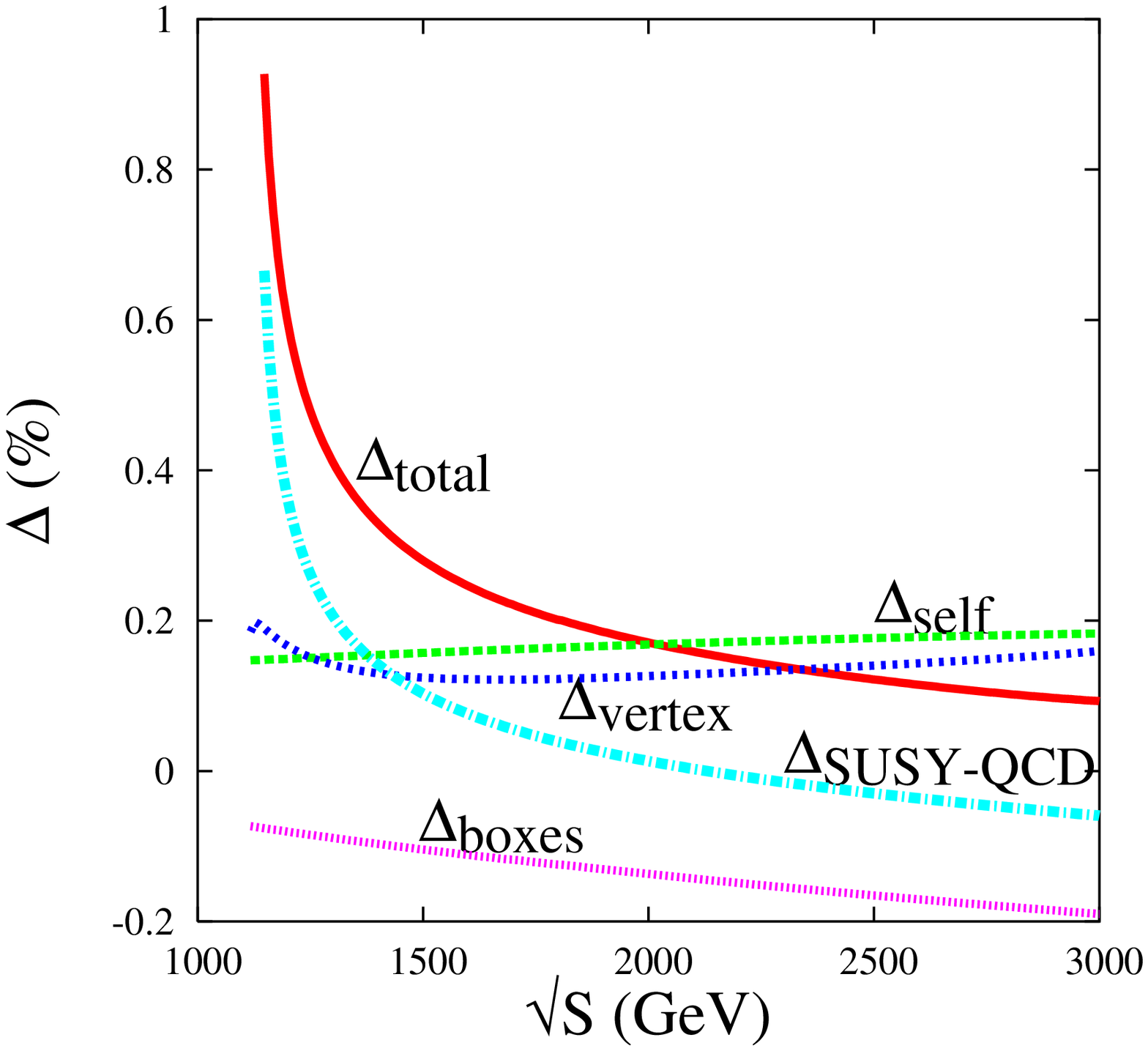} }
\vskip-4.cm
\smallskip\smallskip 
\centerline{{
\epsfxsize2.765 in 
\epsffile{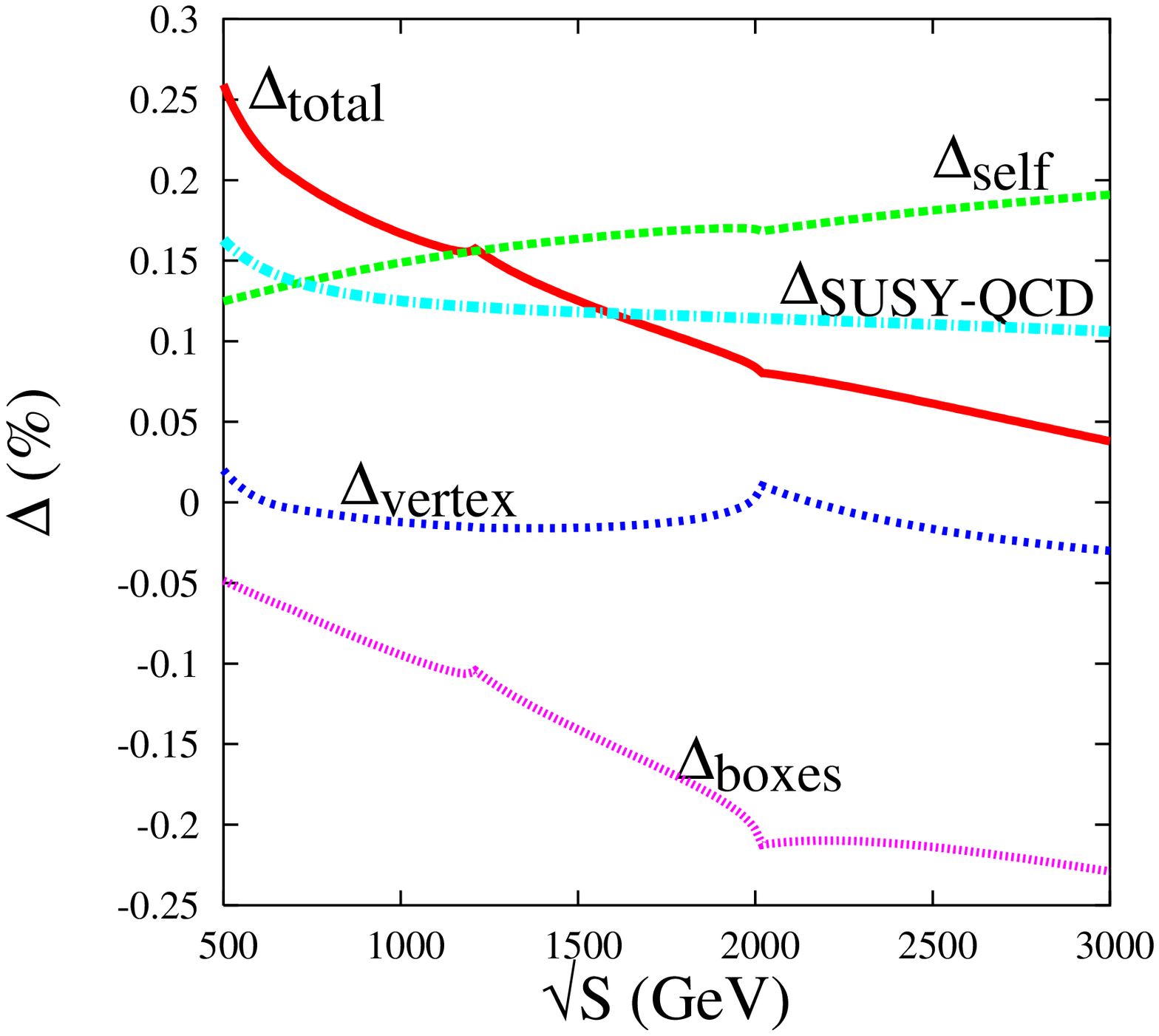}}  \hskip0.4cm
\epsfxsize2.765 in 
\epsffile{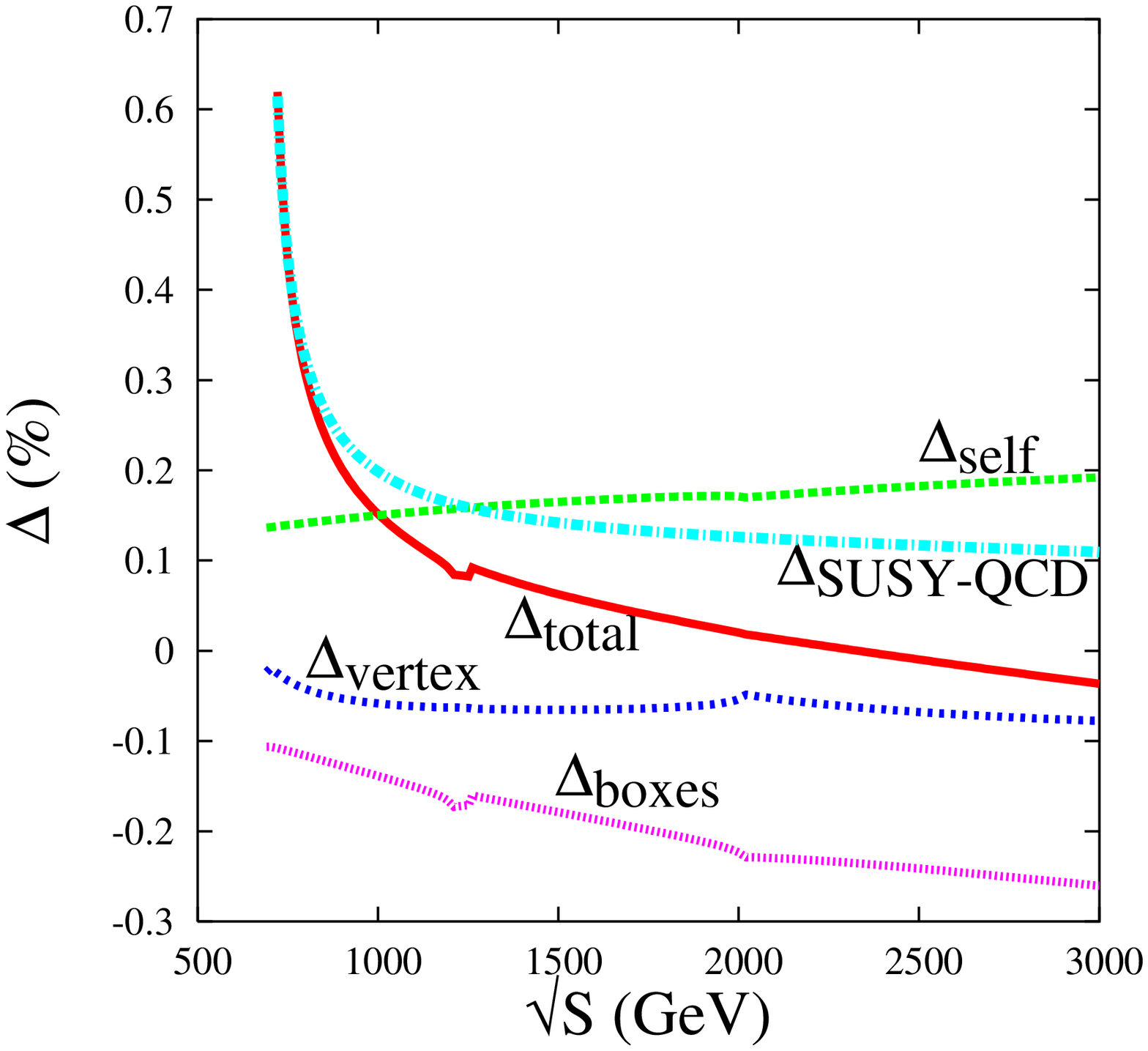} }
\smallskip\smallskip
\caption{Electroweak and SUSY-QCD corrections  to 
$e^+e^- \to \wt{t}_1\wt{t}_1^*$ (left) and $e^+e^- \to \wt{t}_2\wt{t}_2^*$ 
(right).
Scenario $sc_1$ (upper plots), $sc_{2}$ (middle plots), 
and $sc_{3}$ (lower plots).}
\label{del2}
\end{figure}

In order to outline the loop contributions more directly,
we display in Fig.~\ref{del1} ($\tilde{b}$ squarks) 
and Fig.~\ref{del2} ($\tilde{t}$ squarks)
the relative corrections to the cross sections and
show also the breakdown into their various subclasses.

A substantial fraction is  due to the SUSY-QCD corrections, which in general
are dominated by the conventional real- and virtual-gluon contributions. 
One can see in Figs.~\ref{del1},~\ref{del2} the enhancement for 
$\sqrt{s}\to 2 m_{\wt{f}_i}$, which indicates the QCD equivalent of the 
QED Coulomb singularity, studied in~\cite{freitas2,DH}
for slepton-pair production. 
The gluon contributions are always positive and sizeable, and 
the gluino contributions are usually small. Only at very 
high energies the gluino effects can become more important; they are
negative and diminish the gluon-QCD corrections. 
In scenario 2, for example, 
at $\sqrt{s}=3$ TeV, the gluino correction can reach about -8\% 
for $\wt{b}_1 \wt{b}_1^*$, -13\% for 
$ \wt{b}_2 \wt{b}_2^*$, -7\% for $\wt{t}_1 \wt{t}_1^*$,
and -18\% for $\wt{t}_2 \wt{t}_2^*$ final states.

\vspace*{0.2cm}
At very high energies
$\sqrt{s} \gg M_W$, large single and double logarithms of the ratio
$s/M_W^2$ become dominating. 
Self energies  contain only single logarithms, which 
correspond to the evolution of running coupling constants.
Box diagrams contain 
Sudakov logarithms of the type $\log^2(s/M^2)$ (where $M$
is a generic mass of internal particles), which are large and negative.  
The dominant contributions arise
from diagrams with the exchange of standard  gauge
bosons. The leading effect 
at high energies thus comes from the box diagrams
and can reach up to -40\% ($-25\%)$ in case of 
bottom squarks (top squarks).
From Figs.~\ref{del1},~\ref{del2} 
one can see that the large 
box effects interfere destructively with SUSY-QCD and self-energy 
corrections, yielding e.g.\ a total relative correction of about 
-20\% in scenario 2 (1) for $e^+e^-\to \wt{b}_{1(2)}\wt{b}_{1(2)}^*$.

For very high energies, 
higher-order corrections have to be considered as well.
Without entering a two-loop computation,
one get at least some partial information on
the size of the  ${\cal O}(\alpha^2)$ corrections 
from the 
square of the one-loop amplitude in~(\ref{dif}).
As an example, we will give in the Table 1 for the case of 
$e^+e^- \to \wt{b}_1\wt{b}_1^*$  
some numerical values to illustrate the effect
from the square of the one-loop amplitude, of 
${\cal O}(\alpha^2)$. Only the dominating box contributions, 
with gauge-boson exchanges ($D_{6,8}$ with $V_{1,2}=Z,W$) 
are shown. 
The first column is 
the interference term of  ${\cal O}(\alpha)$ [eq.~(\ref{square1})]  
and the second column 
includes also the one-loop square term [eq.~(\ref{square2})]
of  ${\cal O}(\alpha^2)$/
As one can see, at lower energies $\sqrt{s} < 1$ TeV,
the effect is marginal (around 1\%),
while at high energy the cross 
section can be reduced by several per cent.
\begin{table}[t]
\begin{center}
\begin{tabular}{|c|c|c|c|c|c|c|}  \hline
$\sqrt{s}$ & \multicolumn{2}{c|}{500GeV } & \multicolumn{2}{c|}{1.5 TeV} & 
\multicolumn{2}{c|}{3 TeV }\\ 
\hline
$ sc_1$ & -.074    &  -.073    & -.22 &  -.20    &  -.36  & -.32  \\ \hline
$ sc_2$     & -.049  & -.048  & -.17  &  -.16 & -.31    &  -.28 \\ \hline
\end{tabular}
\caption{Relative corrections from the interference term 
 (left column) and with the quadratic one-loop term (right column)
 of eq.~(\ref{dif}) }   
\end{center}
\end{table}

\begin{figure}[t!]
\smallskip\smallskip 
\vskip-4.cm
\centerline{{
\epsfxsize2.8 in 
\epsffile{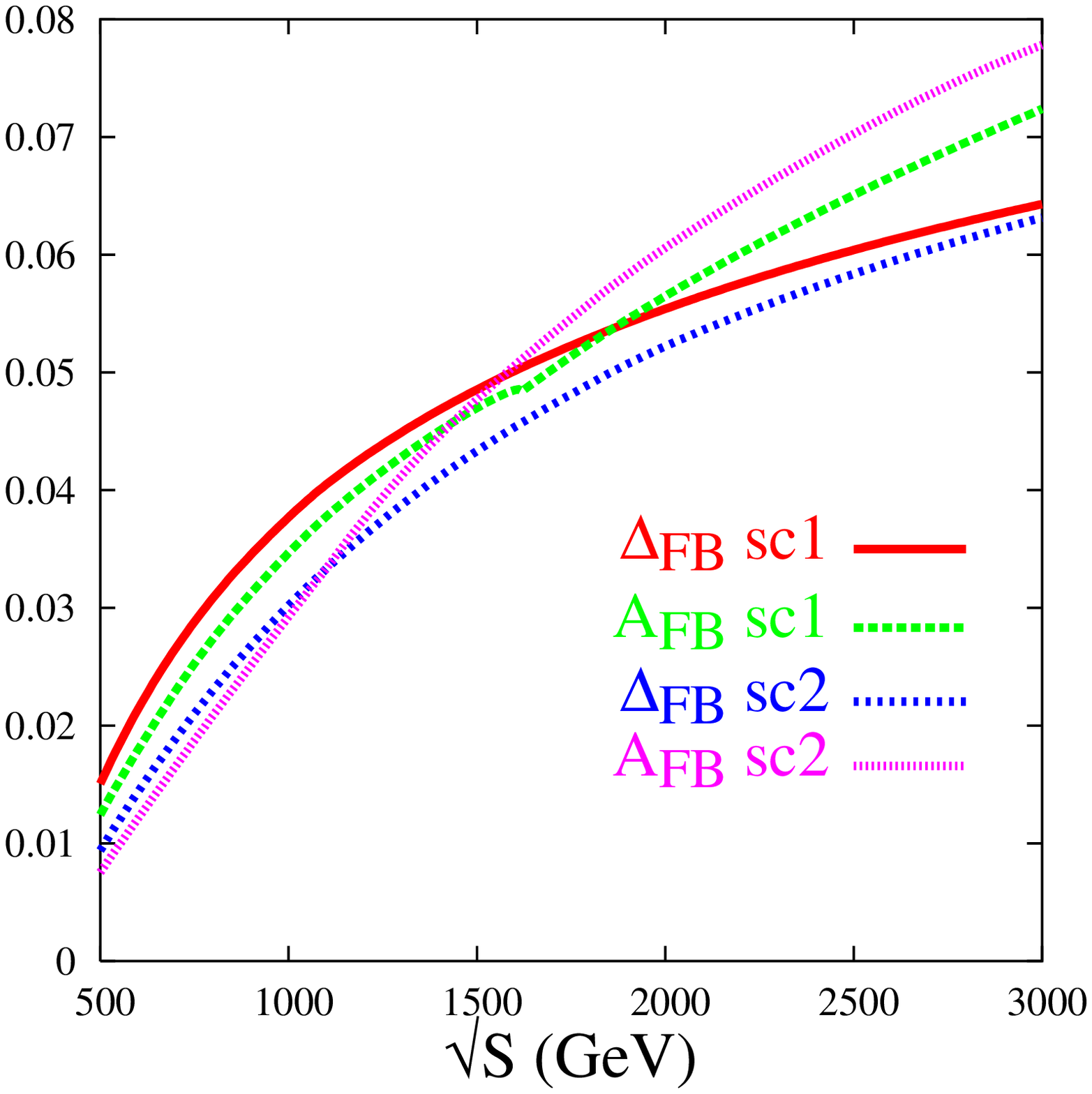}}  \hskip0.4cm
\epsfxsize2.8 in 
\epsffile{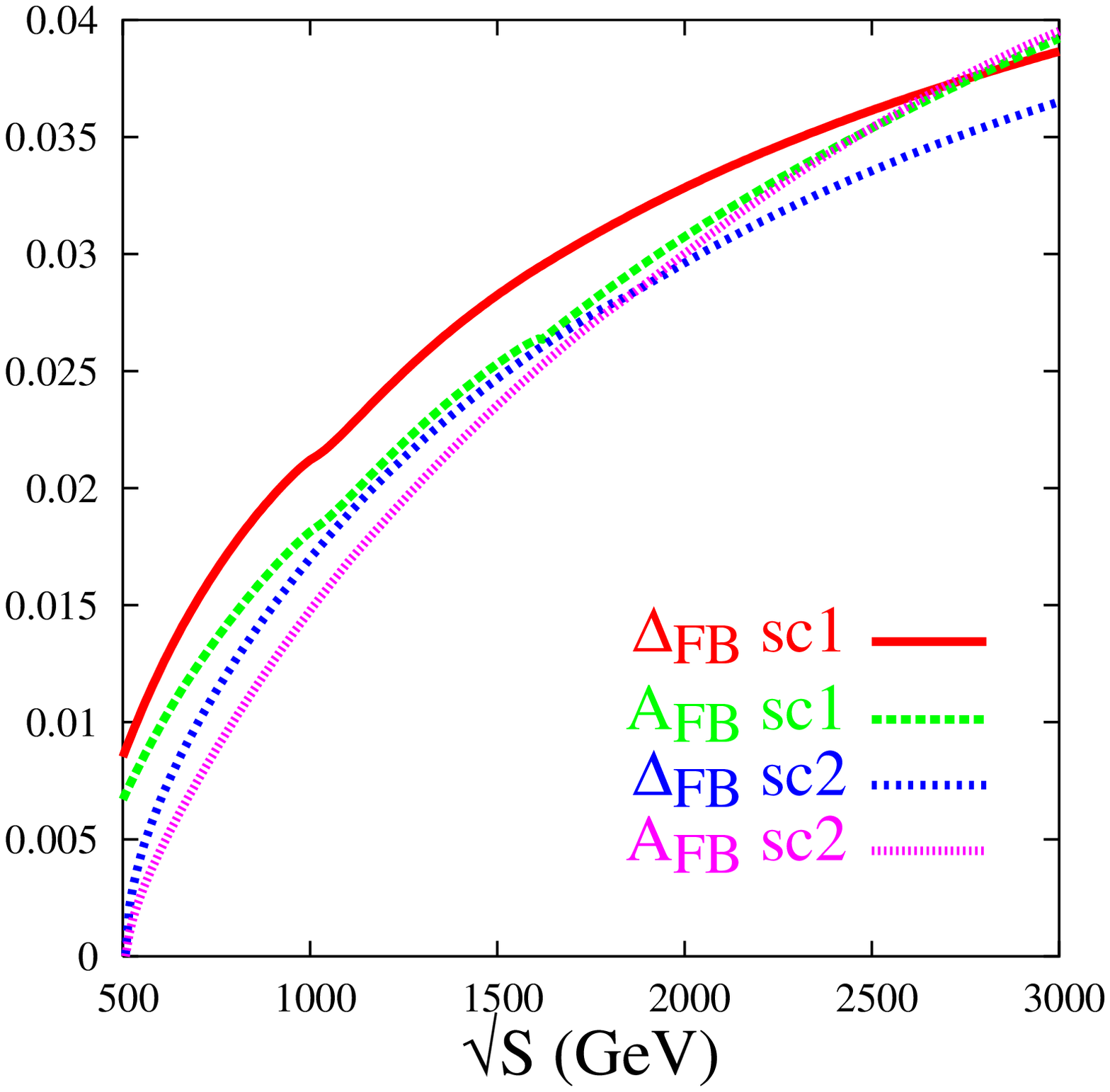} }
\caption{Forward-backward asymmetry  for $e^+e^- \to \wt{b}_1
\wt{b}_1^*$ (left) and $e^+e^- \to \wt{t}_1
\wt{t}_1^*$ (right) as function of center of mass energy 
for scenarios $sc_{1,2}$}
\label{fb}
\end{figure}

In refs.~\cite{claudio1, claudio2}
general one-loop expressions of the leading quadratic 
and subleading linear logarithms of Sudakov type and also 
the corresponding resummation to subleading logarithms are
derived.  
It is difficult to do a quantitative comparison
because the input is not completely specified 
in~\cite{claudio1, claudio2}, but the results are in qualitative 
agreement with ours.

\vspace*{0.2cm}
We want to illustrate also the loop-induced forward-backward asymmetry,
selecting $sc_1$ and $sc_2$ as examples
in Fig.~\ref{fb}.
Contrary to the QED case, 
where only photon and $Z$ are exchanged in  box graphs,
the presence of neutralinos, charginos, and heavy  gauge bosons
in the box diagrams
generates forward-backward asymmetries of several per cent,
about 4\% (7\%)
at 1.5 TeV (3 TeV) for
$e^+e^- \to \wt{b}_1 \wt{b}_1^*$. In the case of 
$e^+e^- \to \wt{t}_1\wt{t}_1^*$, the asymmetry is smaller by 
a factor 1/2.

\begin{figure}[t!]
\smallskip\smallskip 
\vskip-4.cm
\centerline{{
\epsfxsize2.85 in 
\epsffile{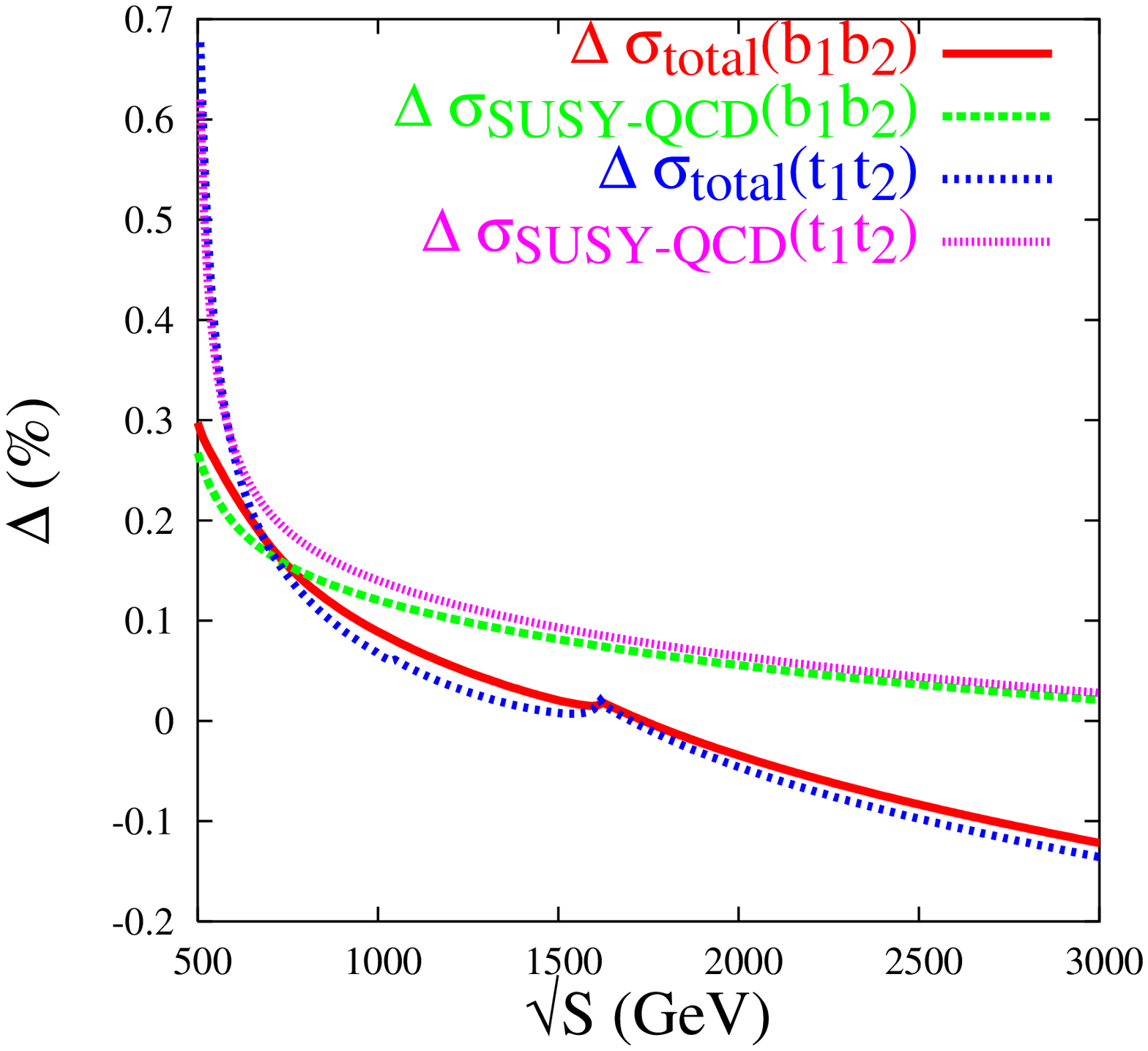}}  \hskip0.4cm
\epsfxsize2.8 in 
\epsffile{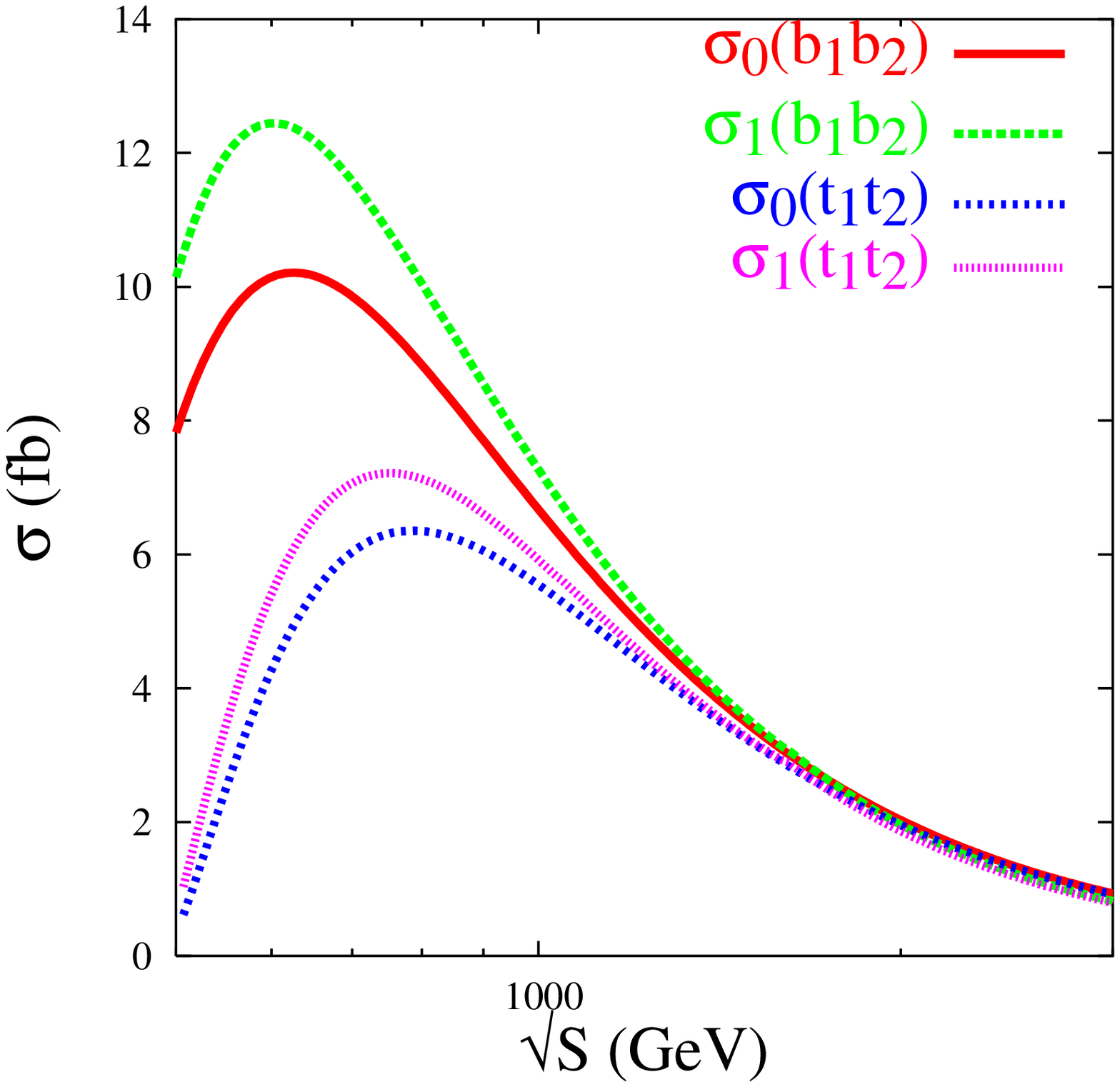} }
\vskip-3.5cm
\centerline{{
\epsfxsize2.85 in 
\epsffile{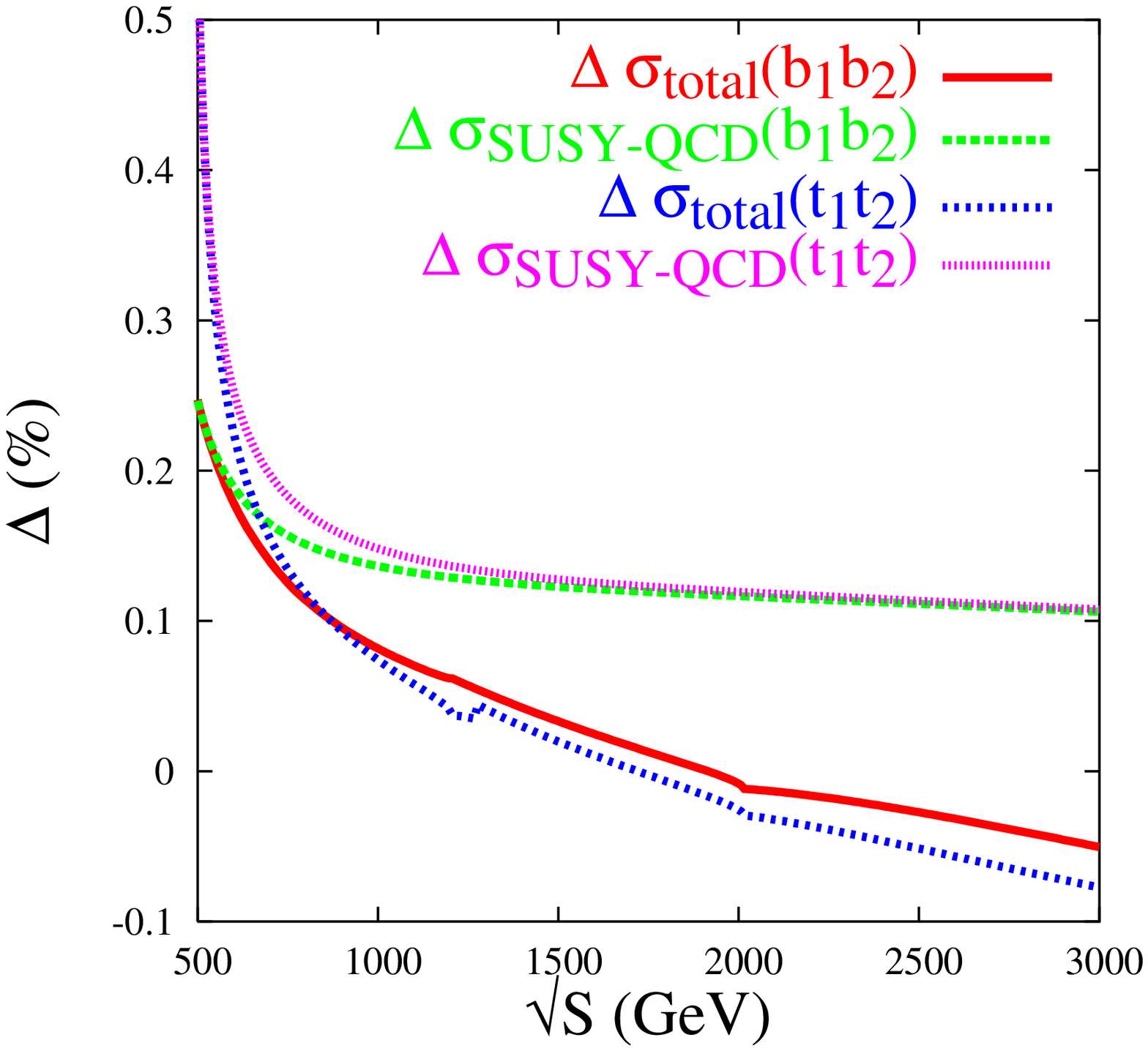}}  \hskip0.4cm
\epsfxsize2.8 in 
\epsffile{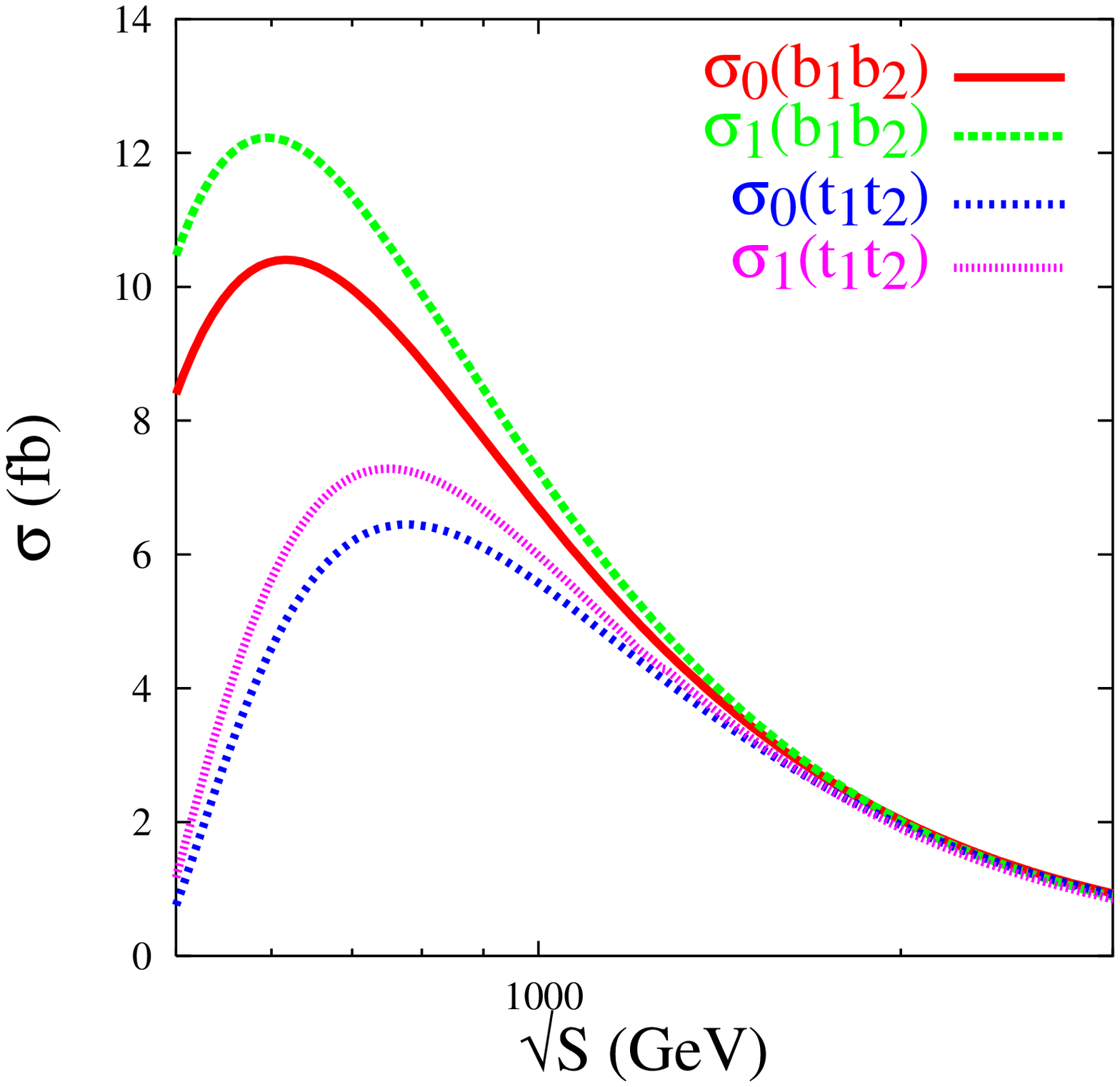} }
\caption{SUSY-QCD and total correction  (left)
and tree-level and one-loop cross section (right) for
$e^+e^- \to ( \wt{t}_1 \wt{t}_2^* , \wt{b}_1 \wt{b}_2^* )$. 
Scenario $sc_1$ (upper plots) and $sc_3$ (lower plots)}
\label{tb12}
\end{figure}

\vspace*{0.2cm}
Finally, we examine the production of 
a non-diagonal squark pair $\wt{f}_1 \wt{f}_2^*$. 
Because of electromagnetic invariance,
the photon does not couple to a pair of $\wt{f}_1$ $\wt{f}_2^*$. 
Consequently, at tree level, 
the reaction $e^+e^- \to \wt{f}_1\wt{f}_2^*$
proceeds only through $s$-channel $Z$-boson exchange. 
The cross section for $e^+e^- \to \wt{f}_1\wt{f}_2^*$ is 
thus directly determined by the 
$Z\wt{f}_1\wt{f}_2^*$ coupling,
which is proportional to
$\sin 2\tilde{\theta}_f$ of the sfermion-mixing angle $\tilde{\theta}_f$.
The measurement of such a process can determine efficiently 
the value of the mixing angle and also the mass of $\wt{f}_2$~\cite{LC}. 
This is complementary to other 
methods proposed to measure the mixing angle, which are 
based on  $e^+e^- \to \wt{t}_1\wt{t}_1^*$
with polarized beams~\cite{LC,Bartl}.
As shown in Fig.~\ref{tb12} for scenarios $sc_1$ and $sc_3$, 
the cross sections for 
$e^+e^- \to \wt{t}_1\wt{t}_2^*$ and $e^+e^- \to \wt{b}_1\wt{b}_2^*$
reach several  femto-barn, 
which can lead to  hundreds of events
for a linear collider with 500 $fb^{-1}$ 
integrated luminosity \cite{LC}.

Without going into details, we mention that the self-energy
contributions are smaller than in the case of diagonal
$\wt{f}_i \wt{f}_i^*$ production, and vertex corrections are more sizeable. 
Box contributions are of about the same size as in the diagonal case 
for $\wt{b}_1\wt{b}_2^*$, 
and are by a factor 2 bigger for $\wt{t}_1\wt{t}_2^*$
compared to $\wt{t}_i\wt{t}_i^*$.

$\ $
\\
Fig.~\ref{tb12} contains also the SUSY-QCD corrections.
One can see that at  lower energies $\sqrt{s}\leq 1$ TeV,
the corrections are dominated by the SUSY-QCD terms.
At higher energies above 1 TeV, the box contributions  
turn out to be sizeable again 
and even dominate, driving the correction negative. 
We note that the forward-backward asymmetry 
for the off-diagonal pair production 
$e^+e^- \to \wt{f}_1\wt{f}_2^*$, also loop-induced, 
is of the same order
as for $e^+e^- \to \wt{b}_1\wt{b}_1^*$. 

\begin{figure}[t!]
\smallskip\smallskip 
\vskip-4.5cm
\centerline{{
\epsfxsize2.85 in 
\epsffile{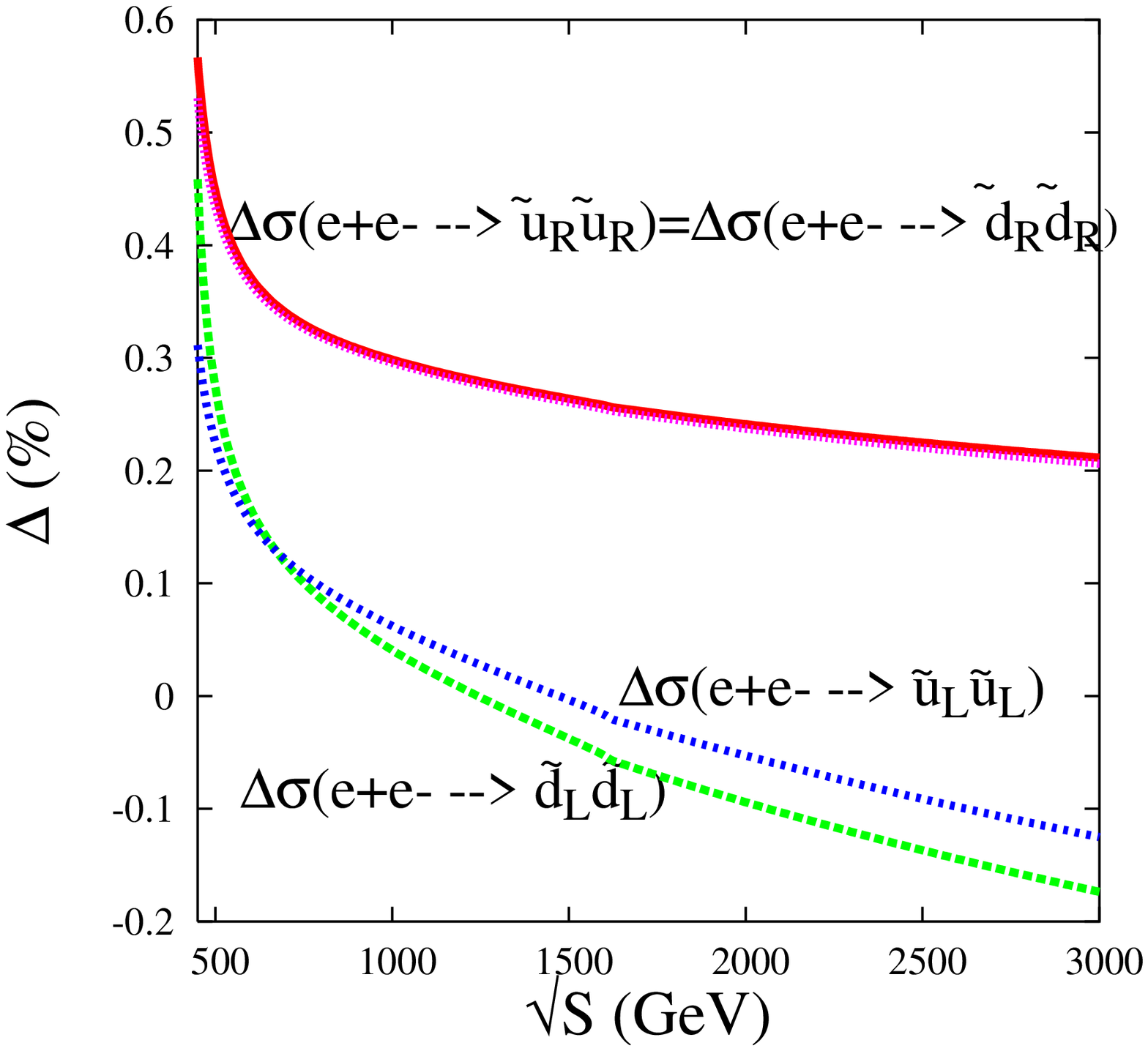}}  \hskip0.4cm
\epsfxsize2.8 in 
\epsffile{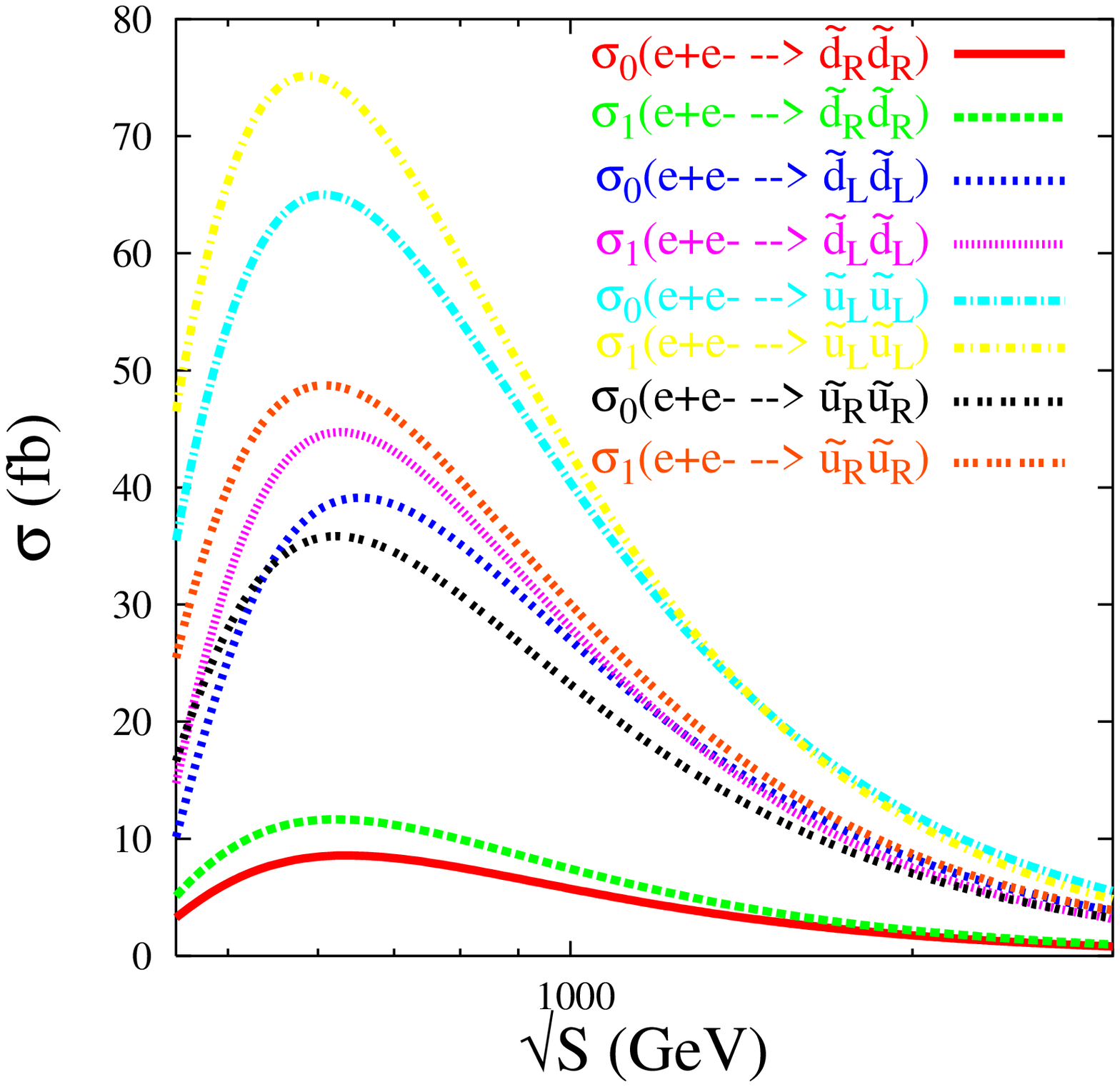} }
\caption{Total relative correction to first- and 
second-generation (up and down) squarks (left) 
and tree-level and one-loop cross 
sections (right), for scenarios $sc_{1,3}$ }
\label{ud12}
\end{figure}

\newpage

\noindent
{\bf First- and second-generation squarks}

We address now the production of squarks 
of the first and second generation.
Since the corresponding quark masses are small,
the squark-mixing angles practically 
vanish, according to eq.~(\ref{mixing}). 
For our scenarios with common $\wt{M}_L = \wt{M}_R$,
we have 
$\wt{d}_{1,2}=\wt{d}_{R,L}$ and $\wt{u}_{1,2}=\wt{u}_{L,R}$.
As an example, in $sc_1$, 
we display in Fig.~\ref{ud12} 
the tree-level and one-loop cross sections together with
the relative correction
for $e^+e^- \to \wt{q}_i\wt{q}_i$, $q=u,d$.
The relative corrections $\Delta(e^+e^- \to \wt{q}_L\wt{q}_L)$ and  
$\Delta(e^+e^- \to \wt{q}_R\wt{q}_R)$ have 
almost the same shape for up and down squarks.
The relative corrections
$\Delta(e^+e^- \to \wt{q}_L\wt{q}_L)$
for the left-handed  squarks are negative at high energies,
while  $\Delta(e^+e^- \to \wt{q}_R\wt{q}_R)$ for the
right-handed squarks are positive. 
This behavior is due to the 
fact that for 
right-handed squarks QCD corrections dominate
and thus remain
positive even at high energy, while for
left-handed squarks box diagrams are dominant being
large and negative at high energy. We recall that
i) the box contribution is dominated by the exchange of 
$W$ and $Z$ bosons, diagrams $D_{6,8}$ of Fig.~\ref{boxes},
ii) the right-handed squarks do not couple to $W$ bosons 
and thus  the box contribution is  small (compared to QCD corrections), 
even at high energy.
\\
\begin{figure}[t!]
\smallskip\smallskip 
\vskip-4.5cm
\centerline{{
\epsfxsize2.6 in 
\epsffile{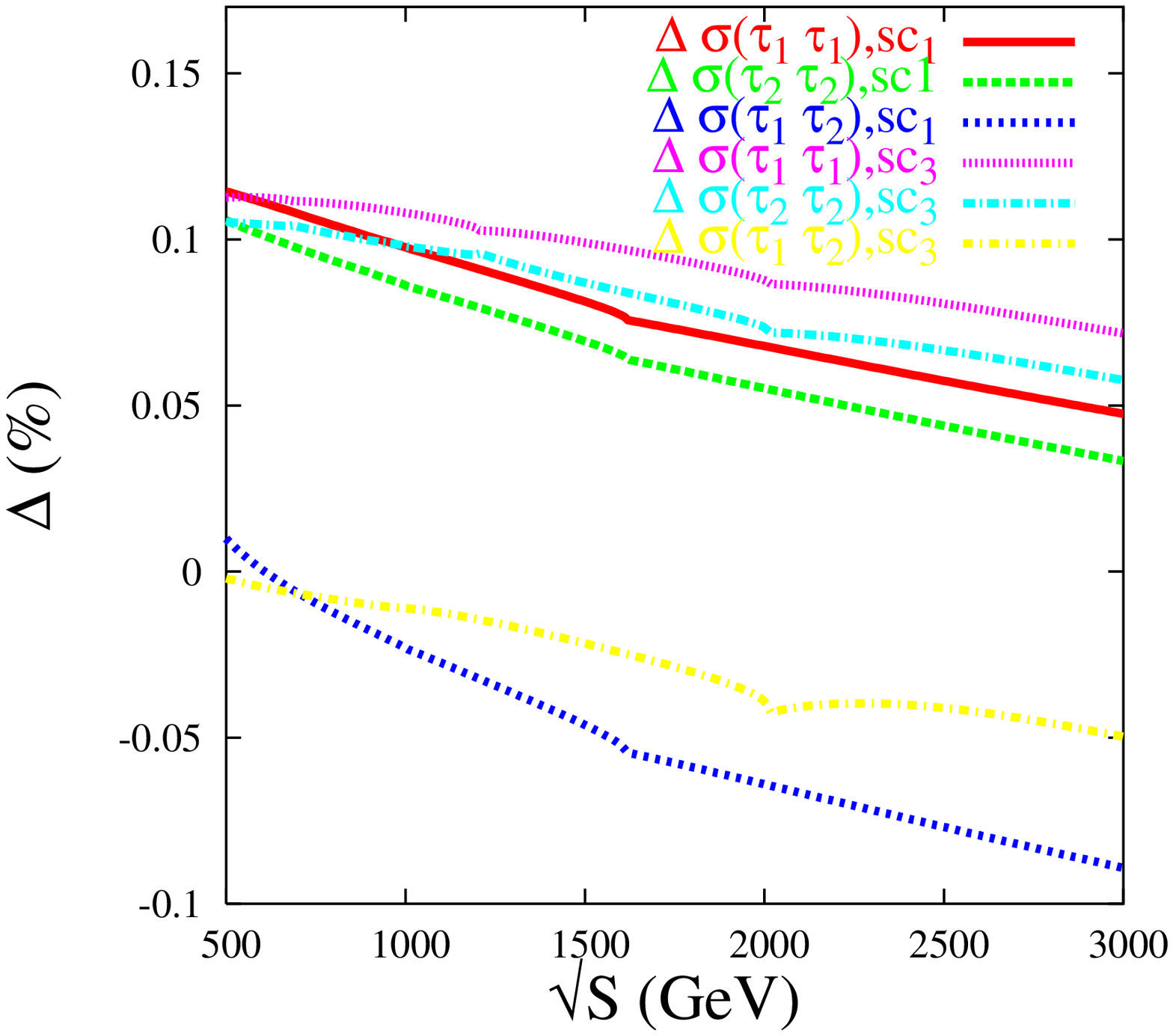}}  \hskip0.4cm
\epsfxsize2.46 in 
\epsffile{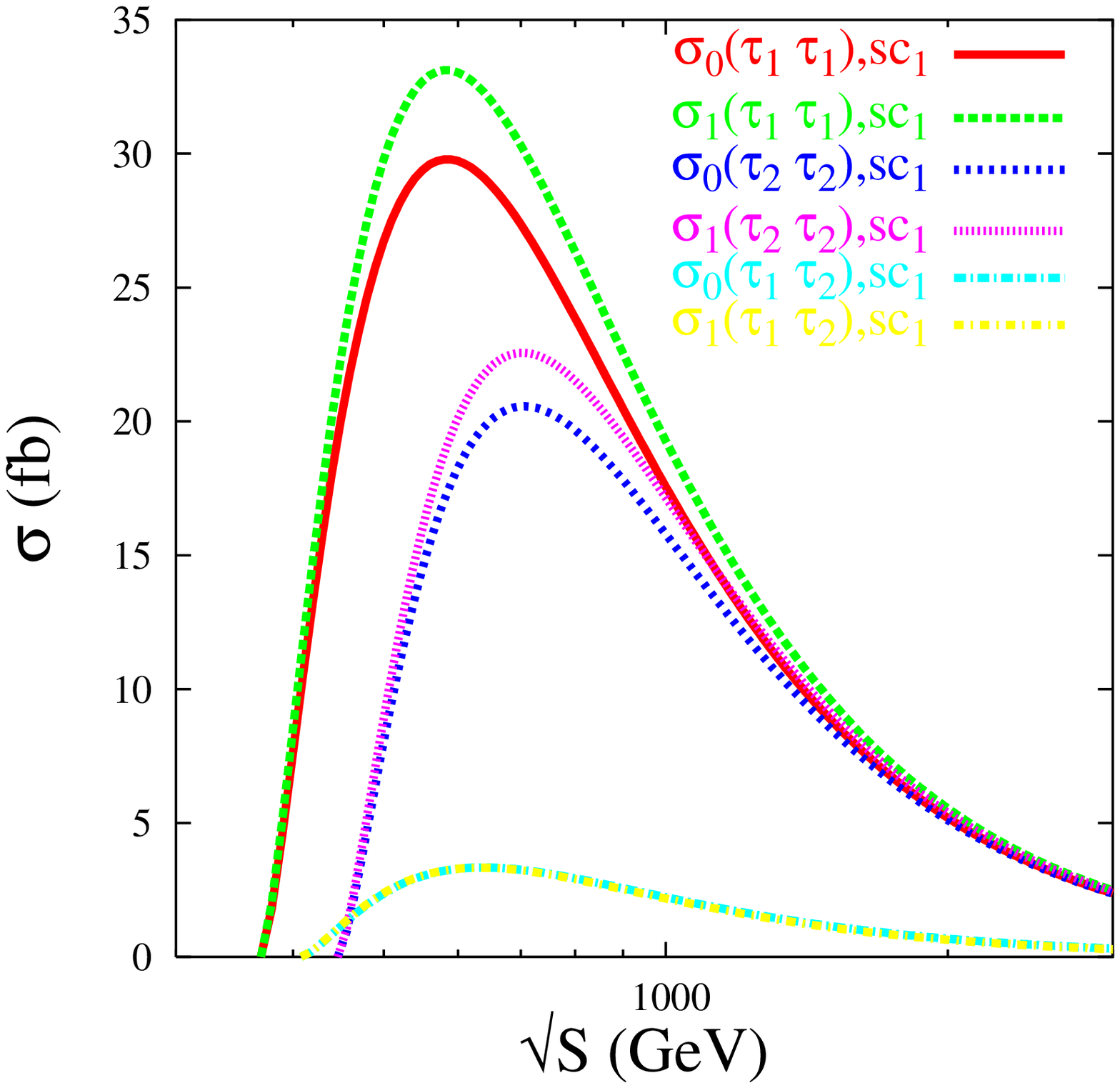} }
\caption{Total relative correction to $e^+e^- \to \wt{\tau}_i
  \wt{\tau}_j^*$ (scenarios $sc_1$ and $sc_2$, left)
and tree-level and one-loop cross sections (scenario $sc_1$, right)
 as functions of the CM energy}
\label{tau}
\end{figure}

\noindent
{\bf $\tau$ sleptons}

In the case of $\wt{\tau}$-pair production, $e^+e^- \to \wt{\tau}_i
\wt{\tau}_j^*$,
shown in Fig.~\ref{tau}, 
the situation is different. There are
no QCD corrections, and the vertex corrections are rather
small, of the order of $\pm 2\%$. The dominant contribution is 
coming from self energies and boxes.
The self-energy contributions are positive in the
three scenarios, while the box contribution is always negative. 
This leads to a destructive interference between boxes and self energies,
keeping the overall corrections comparatively small.
At very high energies,
the self-energy corrections (which get to the order of 20\%) are
reduced by  box and vertex corrections to a net 
effect resulting in  less than 10 \%.
As  illustrated in Fig.~\ref{tau}, the radiative 
corrections to $e^+e^- \to \wt{\tau}_2
\wt{\tau}_2^*$ have almost the same order of magnitude as 
$e^+e^- \to \wt{\tau}_1\wt{\tau}_1^*$
and are positive, while the corrections to 
$e^+e^- \to \wt{\tau}_1 \wt{\tau}_2^*$ turn out to be negative.
This is due to the fact that box corrections to $e^+e^- \to \wt{\tau}_1
\wt{\tau}_2^*$ are the dominant ones.

\section{Conclusion}
We have performed a full one-loop calculation
of scalar-fermion
pair production in $e^+e^-$ annihilation in the on-shell scheme,
comprising electroweak and SUSY-QCD contributions.
We have studied the radiative corrections 
to third-generation scalar fermions $\wt{t},\wt{b}, \wt{\tau}$ 
as well as 
to the first and second generation of scalar quarks.
The QED contributions can be isolated, 
as a separate subclass, and the 
remaining set of one-loop corrections can be
written as the sum of SUSY-QCD, 
self-energies, vertex corrections,  and box contributions.
It has been shown that the self-energy and QCD corrections 
always interfere destructively with the box diagrams.
At low center-of-mass energies, the corrections are dominated by  
QCD, while at high energy the box contributions dominate. 
Box diagrams also induce forward-backward asymmetries
of several per cent.
We have evaluated the radiative corrections to
both diagonal and  off-diagonal pair production; processes like
$e^+e^- \to \wt{f}_1 \wt{f}_2^*$ 
may be useful to extract the mixing angles of the third generation.
In general,
the size of the loop effects, typically of the order 10 -- 20 \%,
makes their proper inclusion in phenomenological studies and analyses
for future $e^+e^-$ colliders indispensable.

\section*{{{Acknowledgment:}}}
A. Arhrib acknowledges the Alexander von Humboldt Foundation.
Part of this work was done within the framework of the 
Associate Scheme of ICTP.  
A. A. thanks ICTP for the warm hospitality extended to him.
We are grateful Thomas Hahn for computing assistance
and useful discussions, as well as to T.~Fritzsche and H.~Rzehak.
We also want to thank H.~Eberl for helpful communications.

\end{document}